\documentclass{emulateapj}
\usepackage[]{natbib, graphicx, float, appendix}
\usepackage[FIGTOPCAP]{subfigure}
\usepackage[section]{placeins}
\usepackage{wrapfig}
\usepackage[fleqn]{amsmath}
\def\vvel{\hbox{${\bf v}$}}

\newcommand{\RNum}[1]{\uppercase\expandafter{\romannumeral #1\relax}}
\newcommand{\uvec}[1]{\boldsymbol{\hat{\textbf{#1}}}}

\begin{document}

\begin{abstract}

Truncated accretion disks are commonly invoked to explain the spectro-temporal variability from accreting black holes in both small systems, i.e. state transitions in galactic black hole binaries (GBHBs), and large systems, i.e. low-luminosity active galactic nuclei (LLAGNs).   In the canonical truncated disk model of moderately low accretion rate systems, gas in the inner region of the accretion disk occupies a hot, radiatively inefficient phase, which leads to a geometrically thick disk, while the gas in the outer region occupies a cooler, radiatively efficient phase that resides in the standard geometrically thin disk.  Observationally, there is strong empirical evidence to support this phenomenological model, but a detailed understanding of the dynamics of truncated disks is lacking.  We present a well-resolved viscous, hydrodynamic simulation that uses an \emph{ad hoc} cooling prescription to drive a thermal instability and, hence, produce the first sustained truncated accretion disk.  With this simulation, we perform a study of the dynamics, angular momentum transport, and energetics of a truncated disk.  We find that time variability introduced by the quasi-periodic transition of gas from efficient cooling to inefficient cooling impacts the evolution of the simulated disk.  A consequence of the thermal instability is that an outflow is launched from the hot/cold gas interface which drives large, sub-Keplerian convective cells in the disk atmosphere.  The convective cells introduce a viscous $\theta-\phi$ stress that is less than the generic $r-\phi$ viscous stress component, but greatly influences the evolution of the disk.  In the truncated disk, we find that the bulk of the accreted gas is in the hot phase.

\end{abstract}


\title{The Dynamics of Truncated Black Hole Accretion Disks \RNum{1}:  Viscous, Hydrodynamic Case}

\author{J.~Drew~Hogg\altaffilmark{1,2,3} and Christopher~S.~Reynolds\altaffilmark{1,3}}

\altaffiltext{1}{Department of Astronomy, University of Maryland, College Park, MD20742}
\altaffiltext{2}{NASA Earth and Space Science Fellow }
\altaffiltext{3}{Joint Space Science Institute (JSI), University of Maryland, College Park, MD20742}

\maketitle

\section{Introduction}
\label{sec-intro}

The radiation produced by the accretion of gas onto a compact object is highly variable, both spectrally and temporally.  One compelling example of this variable behavior is the spectral state changes from outbursting black hole binaries (BHBs) and the associated hysteresis.  BHBs are relatively rare, but the several dozen that have been identified are found to have common evolutionary properties during their state transitions (e.g. Cygnus X-1, 4U 1543-47, H1743-332, GRO J1655-40, GX 339-4).

Much uncertainty surrounds this behavior, but spectral timing studies find that, in a broad sense, there are two primary BHB accretion states, low and high, with distinct properties and additional transitionary intermediate states with composite properties \citep{2006ARA&A..44...49R}.  Of the two main states, the low state is less luminous and the BHB X-ray spectrum is dominated by a power law with index $\sim1.7$.  A thermal spectral component is present, but typically contributes around $20\%$ of the $2-20$ keV flux.  Radio emission originating from a jet is usually observed and, in some objects with high luminosities, an outflowing wind can be found \citep{2016ApJ...830L...5H}, too.  In the high state, a BHB is more luminous.  Thermal, blackbody emission dominates the X-ray spectrum and accounts for $>75\%$ of the unobscured $2-20$ keV flux.  The peak of this thermal emission is around $1$ keV, which is taken as evidence the geometrically thin disk extends close to the inner-most stable circular orbit (ISCO).  In this state, a jet is absent, but signatures of outflowing winds are usually found in the X-ray spectrum.  Additionally, quasiperiodic oscillations (QPOs) are observed during accretion state transitions, with high-frequency QPOs observed in certain intermediate states and low-frequency QPOs found in some low-states.

While these common behaviors exist, every outbursting BHB is unique and has its own peculiarities.  Nevertheless, the gross differences between the low and high state properties seem to naturally be explained with a model where the accretion occurs through either a hot or cold mode.  These different modes are associated with different accretion flow structures and varying degrees of inner disk truncation \citep[for review see][]{2004PThPS.155...99Z, 2006csxs.book..157M, 2007A&ARv..15....1D}.  The preferred mode depends on the cooling efficiency of the inner region of the accretion flow as the accretion rate evolves; specifically, the flow is in a radiatively inefficient mode in the low state and a radiatively efficient mode in the high state.  In the low state, the inefficient cooling in the inner region effectively truncates the cold, thin disk since the thermal energy of the gas cannot be radiated away.  The inner disk is replaced by a hot, diffuse flow, hence, the accretion flow should inflate and be geometrically thick and optically thin.  In the high state, radiation can efficiently remove thermal energy, allowing the disk to take on an optically thick and geometrically thin configuration like the $\alpha$-disk of \citet{1973A&A....24..337S}.  Since the disk can cool efficiently, there is no disk truncation and the thin disk extends down to the ISCO.

The bifurcation of accretion states in active galactic nuclei (AGNs) is less clear, but there is evidence that accretion in low-luminosity AGN (LLAGN), including Sgr A* in our galactic center, is dominated by a hot mode \citep{1982Natur.295...17R, 1995ApJ...444..231N}.  In AGNs whose accretion rate is sufficiently low, less than $1\%$ of their Eddington limit, hot mode accretion \citep{2008ARA&A..46..475H, 2014ARA&A..52..529Y} and inner disk truncation \citep{2011ApJ...733...60T, 2011ApJ...726...87Y, 2014MNRAS.438.2804N} seems to be common.  Qualitatively, the properties of LLAGN resemble the BHB hard state.  The SED of a standard, unobscured, luminous Seyfert ($1-30\%$ Eddington), on the other hand, typically has a large thermal component peaking in the far-UV, analogous to the soft state of BHBs.  The prevalence with which thick-to-thin disk transitions are invoked to explain the phenomenology of both BHBs and AGNs warrants a rigorous investigation of the behavior of the accretion flow at this interface to solidify the theoretical underpinnings of the truncation model.

For gas to accrete onto the central black hole, it must lose angular momentum.  The seminal work of \citet{1973A&A....24..337S} parameterized the strength of the internal stress relative to the gas pressure through the so-called ``$\alpha$-parameter," further described in Section \ref{sec-sim_setup}.  The source of this anomalous viscosity was assumed to be turbulence, but the driving mechanism was not elucidated until the magnetorotational instability (MRI) was proposed as a viable instability \citep{1991ApJ...376..214B}.  The connection between MRI-driven turbulence and the $\alpha$-parameter has since been well-studied in global MHD accretion disk simulations \citep[e.g.][]{1998ApJ...501L.189A, 2000ApJ...528..462H, 2011MNRAS.416..361B, 2011ApJ...735..122F} and thin disks have been shown to behave viscously.

The behavior, energetics, and dynamics of the hot phase accretion, on the other hand, are not well constrained.  Many types of hot flows have been suggested to fill this role including advection dominated accretion flows \citep[ADAFs;][]{1994ApJ...428L..13N, 1995ApJ...444..231N, 1995ApJ...452..710N} and convection dominated accretion flows \citep[CDAFs;][]{2000ApJ...539..809Q}.  Simulations of non radiative, thick disks with varying levels of sophistication have been used to shed light on the behavior of hot accretion flows.  Both HD \citep{1996ApJ...464..364S, 1999MNRAS.303..309I, 2000ApJ...537L..27I, 2003ApJ...582...69P, 2010MNRAS.408.1051Y} and MHD \citep{2001MNRAS.322..461S, 2001ApJ...554L..49H, 2002ApJ...573..738H, 2003ApJ...592.1042I} numerical models ubiquitously develop outflows.  In the HD cases, and some early MHD models, circulation often developed within the outflow and is interpreted as convection.  In more recent MHD models, though, the convection is found to be stabilized \citep[e.g.][]{2012ApJ...761..130Y, 2012MNRAS.426.3241N, 2015ApJ...804..101Y}.

Despite the thoroughness with which numerical models have investigated the thin and thick disk configurations separately, the details of how the hot and cold flow coexist are not understood.  Several global accretion models have been proposed and either explored analytically or with simple numerical models to determine how a truncated disk might develop.  For instance, a truncated disk could exist because the inner disk evaporates due to the cool, Keplerian disk being heated by a hot corona \citep{1994A&A...288..175M, 1996ApJ...457..821N, 1999ApJ...527L..17L, 2000A&A...360.1170R}.  Alternatively, the diffusion of energy from the interior of the disk could also heat the gas beyond the viral temperature, thereby driving the inner disk to a hot accretion solution and truncating the disk \citep{1996PASJ...48...77H, 2000ApJ...529..127M, 2000ApJ...538..295M}.  Local thermal instability can also develop if the $\alpha$-parameter is proportional to the magnetic Prandtl number \citep{2014MNRAS.441..681P, 2017arXiv170402485P}.  However, there is still much uncertainty underlying the trigger mechanism.

Still, headway can be made in characterizing the flow dynamics of a truncated disk.  Of the outstanding issues, there is a particular need to understand the structure of the transition from the assumed thin disk to the thick disk and the influence the truncation has on the disk evolution.  For instance, how does the flow conserve mass, angular momentum, and energy when it passes through an abrupt transition zone?  Can a steady-state configuration develop, or is time-variability a necessary ingredient?  On these fronts, numerical models provide a valuable tool and way to glean insight into the properties of truncated disks.

Radiation HD \citep[RHD;][]{2013MNRAS.435.2431D, 2016MNRAS.459.1543W} simulations have tried to directly address the viability of the state transitions of BHBs.  They find that when realistic radiation prescriptions which account for bremsstrahlung, synchrotron, and synchrotron self-Comptionization are included in the models, disks that are qualitatively similar to the BHB state transition model readily develop.  These models naturally form truncated, thin, cold disks embedded in a hot, diffuse medium.  Recently, \citet{2016ApJ...826...23T} used a simulation which accounted for general relativity, radiation and MHD (GRRMHD) and found a truncated disk developed where gas resided in a two-phase medium configuration.  In the inner region of their simulations, the gas is radiatively inefficient and the viscous time is shorter than the cooling time.  While it was not explicitly stated, this presumably sets up an ADAF-like flow.  Evidence for thermal, and possibly viscous instability, has also been found in the GRRMHD simulations in \citet{2016MNRAS.463.3437M}.

The truncated disks found in the RHD and GRRMHD simulations suggest these transitions can naturally happen, but relying on computationally intensive algorithms comes at the sacrifice of resolution and simulation time.  A pervasive issue in these works is that the cold disk cools to the scale of the grid and the internal dynamics are unresolved.  This poses a problem because the defining evolution of the accretion flow can occur on small scales and on long times, which makes a proper simulation of a truncated accretion disk very difficult to do with current computational capabilities.  Thus, to conduct a detailed study of the accretion flow itself and get a handle on how a truncated accretion disk will evolve, we must make the problem tractable by forgoing some of the more computationally expensive physics, like a full treatment of the radiation physics and the general relativistic effects which may impact the evolution of the disk in the very inner region close to the black hole.  Here, we assume the truncation found in the simulations of \citet{2013MNRAS.435.2431D, 2016MNRAS.459.1543W, 2016ApJ...826...23T, 2016MNRAS.463.3437M} can naturally occur and seek to fully characterize the accretion flow of a truncated $\alpha$-type viscous disk and understand the observational consequences.  

In this paper we produce the first sustained truncated accretion disk, i.e. a truncation that can be studied for thousands of dynamical times.  Our simulation is well-resolved and allows for a detailed study of the dynamics, energetics, and angular momentum transport in the disk.  We do not use a full prescription for the radiation, but rather use an \emph{ad hoc}, bistable cooling function to drive a thermal instability interior to some radius, thereby creating a transition between a cold and hot disk that can be characterized.  \emph{We emphasize that the aim of this work is not to reproduce the state transitions observed in BH systems, instead it is to study the putative truncated disk found in hard state BHBs and LLAGNs.}  To isolate the behavior of the transition zone, we study the most ideal scenario possible, an axisymmetric viscous HD accretion disk in a Newtonian potential.  The work presented here (Paper \RNum{1}) is a precursor to a study using a 3D high-resolution MHD simulation with the same bi-stable cooling function (Paper \RNum{2}).  Together, these investigations will significantly improve our understanding of truncated disk dynamics.

The viscous formalism used here gives the freedom to set the viscosity and balance the heating rate to the cooling rate according to an $\alpha$-parameter.  To better match the accompanying MHD truncated disk simulation, we set $\alpha=0.065$.  This value was determined through a calibration MHD simulation, described in Paper \RNum{2}.  In setting the HD $\alpha$-parameter to that of an effective $\alpha$-parameter from an MHD model, we can provide a more direct association of the disk behaviors between the HD and MHD regime.

By first establishing a benchmark with a simple $\alpha$-disk, we hope to gain a better understanding of the more complex disk model set in MHD, where heating and angular momentum transport occur solely thorough MRI-driven turbulence, and more honestly assess the viability of truncated disk models in the literature.  Much of the previous theoretical work is developed in HD, so the simplified model considered here will help bridge the gap between expectations of models developed in the literature and the more MHD model.

This paper is organized as follows.  In Section \ref{sec-nm} we discuss our numerical model, simulation setup, and bistable cooling function.  In Section \ref{sec-res} we discuss our analysis of dynamics, angular momentum transport, and energetics of the truncated disk.  In Section \ref{sec-disc} we discuss the implications of our results and the caveats associated with the simulation.  In Section \ref{sec-conc} we provide a summary and closing remarks.

\section{Numerical Model}
\label{sec-nm}

The goal of our work is to emulate a truncated accretion disk and perform a detailed study of its dynamics, angular momentum transport, and energetics.  We therefore idealize our test problem so that we can direct our computational resources to the physics necessary to capture the hot-cold flow boundaries, run the model at high resolution, and run for many evolutionary times.  In this work, we consider a two-dimensional, axisymmetric, viscous disk.  The simulation is divided into two phases: an initialization period before the switch to the bistable cooling function and the latter portion used in our analysis where the bistable cooling function drives thermal instability and creates a sustained truncation.  For numerical reasons that become especially important in the MHD case, we initially evolve the simulation as a thin disk with constant aspect ratio $h/r=0.1$ where the disk thickness is maintained by a cooling function with a single target disk thickness.  Once a quasi-steady state develops, the cooling is switched to a bistable cooling function where the gas can cool efficiently with a target scale height $h/r=0.1$ or inefficiently with a target scale height of $h/r=0.4$, depending on its temperature.  If the gas temperature is above the transition temperature, $T_{T}$, it cools according to the inefficient prescription; if it is below it cools according the the efficient prescription.  The analysis focuses on the truncated disk.  

The additional computational expense of accounting for full GR in this hydrodynamic simulation is manageable; however, it would be very restrictive for our follow-up simulation in MHD and hinder our ability to achieve the goal of performing a detailed study of the flow behavior.  Therefore, we opt to consider the simplified Newtonian case here to aid in comparing between the HD and MHD scenarios.   Additionally, these transitions are expected to occur at $30-100$ $r_g$ \citep{2014MNRAS.438.2804N}, which is far enough from the black hole that relativistic effects are minimal, so there is observational motivation for neglecting GR as well.  

\subsection{Simulation Code}

For this study we use the second-order accurate PLUTO \emph{v4.2} code \citep{2007ApJS..170..228M} to model the gas dynamics.  PLUTO employs a Godunov scheme that evolves the equations of hydrodynamics, \begin{eqnarray}
\frac{\partial \rho}{\partial t}+\nabla\cdot(\rho\vvel)&=&0,\\
\frac{\partial}{\partial t}(\rho\vvel)+\nabla\cdot(\rho\vvel\vvel+P{\cal I})&=&-\rho\nabla\Phi+\nabla\cdot\mathbf{\sigma},\\
\frac{\partial}{\partial t}(E+\rho\Phi)+\nabla\cdot\left[(E+P+\rho\Phi)\vvel\right]&=&\nabla\cdot(\mathbf{\sigma}:\vvel)-\Lambda,\\
\nonumber
\end{eqnarray}
where $\rho$ is the gas density, $\vvel$ is the fluid velocity, $P$ is the gas pressure, ${\cal I}$ is the unit rank-two tensor, {\boldmath$ \sigma$} is the viscous stress tensor, $E$ is the total energy density of the fluid,
\begin{equation}
E=u+{1\over 2}\rho |\vvel|^2,
\end{equation}  
and $\Lambda$ accounts for radiative losses through cooling.  PLUTO was used in the dimensionally-unsplit mode with the {\tt hllc} Riemann solver (a three-wave, four state solver) with piecewise linear spatial interpolation and second-order Runge-Kutta time integration.  The Godunov scheme is a finite-volume method which conserves the total energy of simulation to machine accuracy.  The only energy losses come from outflow across the domain boundary and the cooling function of the simulation.

\subsection{Simulation Setup}
\label{sec-sim_setup}

In our simulation we adopt an axisymmetric, two-dimensional, spherical coordinate system ($r$, $\theta$, $\phi$) for our computational grid.  The domain spans $r \in [10 r_{g}, 1000 r_{g}]$ and $\theta \in [\pi/2-1.0, \pi/2+1.0]$ with $N_R \times N_\theta  = 512 \times 512 = 2.62 \times 10^5$ zones.  The radial spacing of the grid increases logarithmically so that $\Delta r / r$ is constant.  In the $\theta$-direction, the simulation uses 320 uniformly grid zones in the midplane region from $\theta=\pi/2\pm0.5$ and 96 geometrically stretched grid zones in each of the coronal upper regions ($\pi/2+0.5 \leq \lvert \theta \rvert \leq \pi/2+1.0$).

Outflowing boundary conditions are used for the inner and outer $r$-boundaries and the $\pm\theta$ boundaries, but were modified to prevent the inflow of material.  Periodic boundary conditions are used for the axisymmetric $\phi$-direction.  A density and pressure floor is implemented to protect against values becoming artificially small or negative.  All fluid variables (e.g. $\rho$, $p$, $T$) are reported in normalized units from the simulation.

We consider an ideal gas with a $\gamma = 5/3$ adiabatic equation of state in a Newtonian gravitational potential where,  \begin{equation}
\Phi=-\frac{GM}{r}.
\end{equation}  Initially, the simulated disk is thin ($h/r=0.1$) and in vertical hydrostatic equilibrium.  The steady state $\alpha$-disk solution is used for initialization: \begin{equation}
\rho(R, \theta)=\rho_{0} R^{-3/2}\exp\Bigg(-\frac{z^2}{2 c_{s}^2 R^3}\Bigg)\Bigg[ 1-\Bigg(\frac{R_{*}}{R}\Bigg)^{\frac{1}{2}}\Bigg]
\end{equation} where $\rho_{0}$ is a normalization constant to set the maximum disk density to unity, $R$ is the cylindrical radius ($R=r \cos\theta$), $z$ is the vertical disk height ($z=r\sin\theta$), and $c_{s}$ is the sound speed ($c_{s}=\sqrt{P/\rho}=(h/r) v_{K}$).  $R_{*}$ is taken to be just inside the inner simulation domain ($R_{*}=11\:GM/c^2$) to account for the torque-free inner boundary condition.

In astrophysical accretion disks, the effective viscosity originates from MRI driven turbulence.  In our hydrodynamic simulation, however, we approximate this by treating the angular momentum transport with an $\alpha$-type viscosity according to \citet{1973A&A....24..337S}.  We set the shear viscosity term to be,
\begin{equation}
\label{eqn-visc}
\nu = \frac{2}{3} \alpha P \Omega_K^{-1} = \frac{2}{3}\alpha \rho T \Omega^{-1}_K,
\end{equation}  where $\Omega_K$ is the Keplerian angular velocity. A dimensionless $\alpha$-parameter is set to $\alpha=0.065$ to match the test MHD run discussed in Paper \RNum{2}, as we described in Section \ref{sec-intro}.  

\subsection{Cooling Function}
\label{sec-cooling_func}

Our goal is to construct a model accretion disk that undergoes a radial transition from a geometrically-thin radiatively-efficient state to a geometrically-thick inefficient state as the matter flows inwards.  Once such a state has been constructed, we can then characterize the dynamics of this transition region.  Here, we describe the toy optically-thin cooling function that we use to achieve this goal.

The volumetric heating due to the $r-\phi$ shear stresses is
\begin{equation}
{\cal H}={3\over 2}\alpha{\cal R}\rho T\Omega_K,
\end{equation}
where $\Omega_K$ is the Keplerian angular velocity, and ${\cal R}$ is the gas constant.  In a steady-state, radiatively efficient disk we will have ${\cal H}=\Lambda$.  We seek to construct a cooling function $\Lambda(\rho,T,r)$ that (i) is positive-definite everywhere (in contrast to many of the ``target thickness" cooling functions employed in previous work; \citealt{2009ApJ...692..411N}), (ii) yields a disk model that is viscously stable, (iii) produces a hot, geometrically-thick solution at small radii, (iv) produces a cool, geometrically-thin solution at large radii.  We adopt a cooling function of the form,
\begin{equation}
\Lambda=\Lambda_0(T)\rho T^2f(r),
\end{equation}
where $f(r)$ shall be specified later to maximize convenience. The linear dependence on density ensures that the thermal balance is independent of density, removing the possibility of viscous instability.  Noting that $c_s^2={\cal R}T=(h/r)^2\Omega_K^2$, we see that the condition ${\cal H}=\Lambda$ gives
\begin{equation}
\frac{h}{r}=\sqrt{\frac{3\alpha{\cal R}^2}{2\Lambda_0(T)r^2\Omega_Kf(r)}}.
\end{equation}
We choose $f(r)={\cal R}^2r^{-2}\Omega_K^{-1} \propto r^{-1/2}$ so that
\begin{equation}
\frac{h}{r}=\sqrt{\frac{3\alpha}{2\Lambda_0(T)}},
\end{equation}
and, by construction, the disk has a constant aspect ratio (i.e., fixed $h/r$) if $\Lambda_0$ is a constant.

In this work, we induce thermal instability and hence a radial transition into the flow by introducing an abrupt decrease in $\Lambda_0(T)$ at a prescribed transition temperature $T_T$, 
\begin{eqnarray}
\Lambda_0(T)=\begin{cases}
\Lambda_c \hspace{1cm}(T<T_T)\\
\Lambda_h \hspace{1cm}(T\ge T_T)\\
\end{cases}
\end{eqnarray}
where $\Lambda_{c}$ and $\Lambda_{h}$ are constants, and we will assume here that $\Lambda_c>\Lambda_h$.  With $\Lambda_0$ so defined, the thermal balance condition ${\cal H}=\Lambda$ delineates three radial regions within the disk.  Inside of a radius $r_{tr,i}=3\alpha GM/2\Lambda_c{\cal R}^2T_T$ (the inner transition radius), local thermal balance requires that the flow is hot with geometric thickness
\begin{equation}
\left(\frac{h}{r}\right)_{\rm hot}=\sqrt{\frac{3\alpha}{2\Lambda_h}}\hspace{0.5cm}{\rm for}\hspace{0.5cm}r<r_{tr,i}:=\frac{3\alpha GM}{\Lambda_c{\cal R}^2T_T}
\vspace{0.01cm}
\end{equation}
Outside of a radius $r_{tr,o}=3\alpha GM/2\Lambda_h{\cal R}^2T_T$ (the outer transition radius), thermal balance requires that the flow is cold with geometric thickness
\begin{equation}
\left(\frac{h}{r}\right)_{\rm cold}=\sqrt{\frac{3\alpha}{2\Lambda_c}}\hspace{0.5cm}{\rm for}\hspace{0.5cm}r>r_{tr,o}:=\frac{3\alpha GM}{\Lambda_h{\cal R}^2T_T}
\end{equation}
Between these two transition radii, the flow is bistable and thermal balance can be achieved either in the cold or the hot branch with geometric thickness
\begin{equation}
\left(\frac{h}{r}\right)_{\rm cold} \hspace{0.2cm}{\rm or}\hspace{0.3cm}\left(\frac{h}{r}\right)_{\rm hot} \hspace{0.3cm}{\rm for}\hspace{0.3cm}r_{tr,i}<r<r_{tr,o}.
\end{equation}
Note that, due to the form of our cooling law, the ratio of the cold/hot geometric thickness is related to the ratio of the inner/outer transition radii,
\begin{equation}
\frac{r_{tr,o}}{r_{tr,i}}=\left[\frac{(h/r)_{\rm hot}}{(h/r)_{\rm cold}}\right]^2
\end{equation}

In this work, we set
\begin{eqnarray}
&\left(\frac{h}{r}\right)&_{\rm cold}=0.1\\
&\left(\frac{h}{r}\right)&_{\rm hot}=0.4\\
&r_{tr,i}&=100r_g.
\end{eqnarray}
which implies $r_{tr,o}=1600r_g$.  As we shall see, this toy cooling function results in a successful transition from a cold/thin to a hot/thicker disk close to the inner transition radius. Interestingly, the flow in the bistable region remains predominantly in the cold/thin state.

\section{Results}
\label{sec-res}

We will now discuss our analysis of the hydrodynamic disk simulation.  We show that a truncated disk does indeed develop and investigate the dynamics, energetics, and angular momentum transport associated with the accretion flow in the truncated state.  Interestingly, we find that a sustained stream of sub-Keplarian gas flows inwards from the thick-to-thin disk transition and undergoes delayed heating.  As the gas heats to $T_{T}$ and beyond, pockets of hot gas form in the dense disk which buoyantly rise and cool, inducing a circular flow in the coronal regions of the disk and CDAF-like behavior.  As the disk refills at the transition radius, the cycle repeats quasi-periodically.

\subsection{The Truncated Disk}
\label{sec-trunc}

\begin{figure*}
\includegraphics[width=\textwidth]{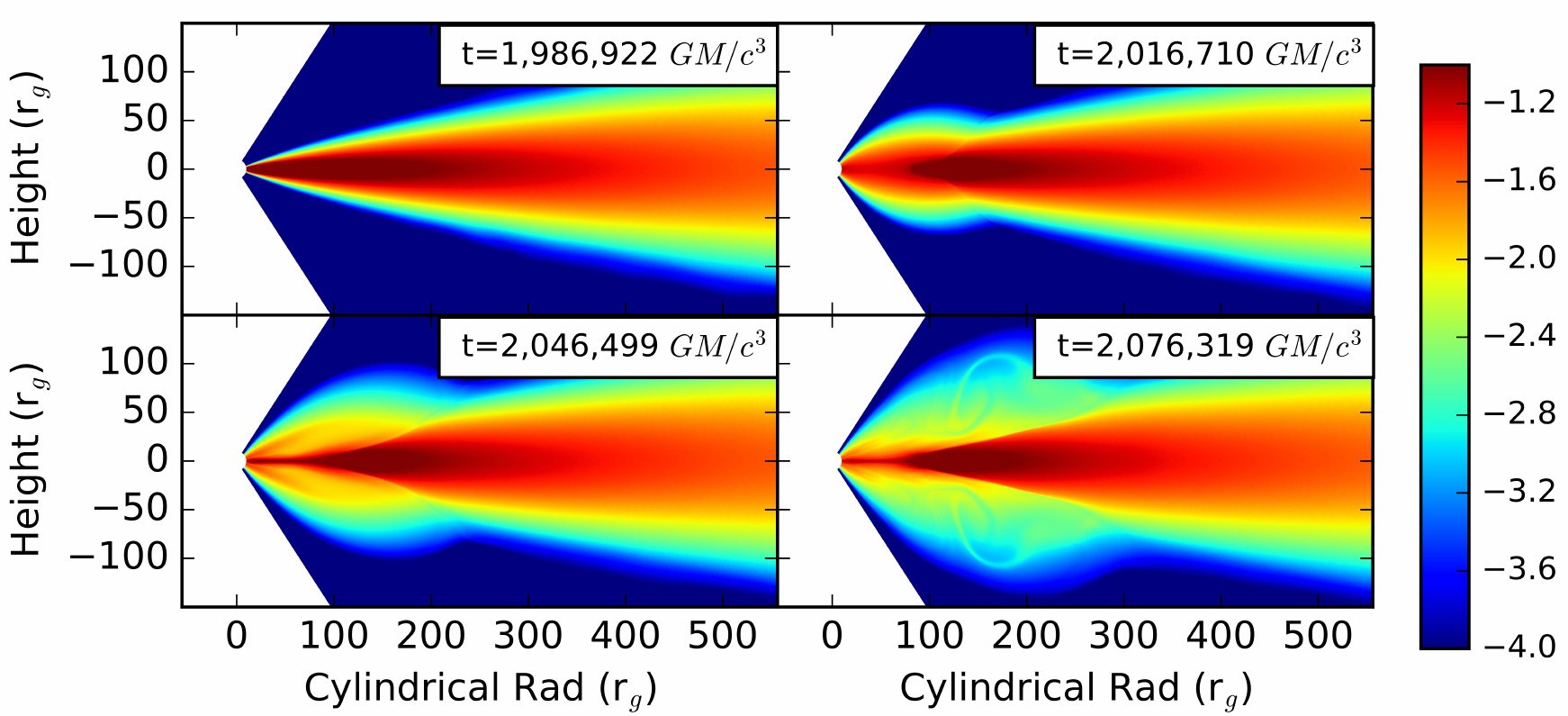}
\caption{Snapshots of the gas density during transient heating event at the end of the disk initialization.  The top left panel shows the simulation when the cooling is switched to the bistable cooling function at $t=1.99\times10^6\:GM/c^3$ ($10,000th$ inner orbit).  The following three panels show the simulation evolution at a $3.0\times10^4\:GM/c^3$ (150 inner orbit times) interval.
\label{fig-trans_4_panel}}
\end{figure*}

We begin by discussing the initial development of the truncated disk in the final phase of the initialization and the general properties of the truncated disk.  As discussed in Section \ref{sec-cooling_func}, the simulation is first evolved with constant $h/r=0.1$, from $t=0$ $GM/c^3$ to $t=1.98\times10^6$ $GM/c^3$ (the first $10,000$ inner orbits) before the cooling is switched to include both a hot and a cold branch.  Immediately after the switch, there is a large, transient heating event within $r\approx200\:r_{g}$ that lasts roughly $\Delta t =1.4\times10^4$ $GM/c^3$ and then cools over $\Delta t = 8.6\times10^4\:GM/c^3$.  At $100$ $r_{g}$, $t_{dyn}\sim6.3\times10^3 GM/c^3$ so the heating event lasts approximately 2.5 dynamical times and the cooling event occurrs on roughly the thermal time of the inner disk.  Figure \ref{fig-trans_4_panel} shows four snapshots of $\rho$ during the heating event.

\begin{figure*}
\centering
\includegraphics[width=0.9\textwidth]{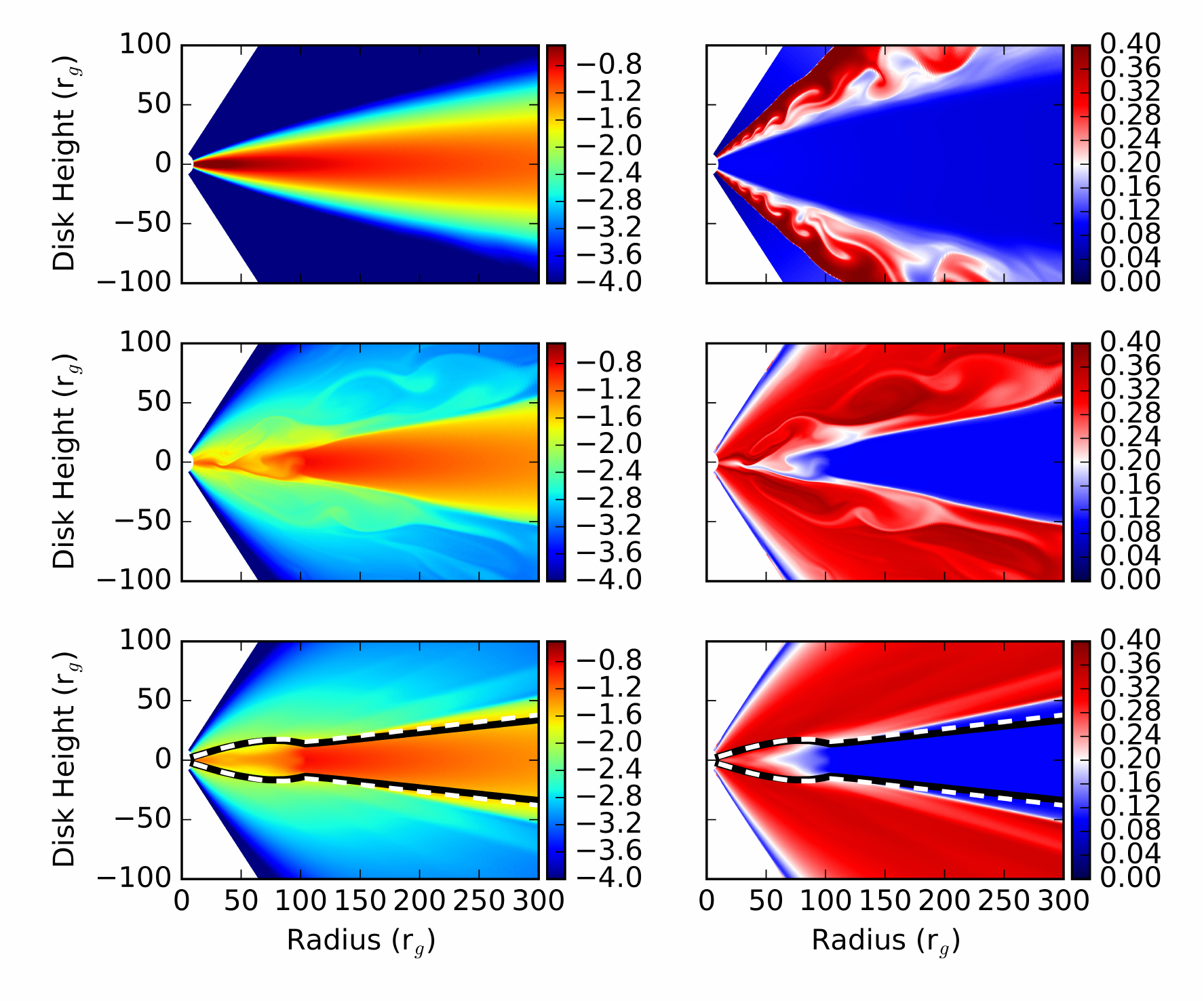}
\caption{Density and temperature structure of the disk.  In the top two rows, the logarithm of the instantaneous density (left column) and temperature scale height (right column) are shown of the inner simulation region during the initialization at the switch to the bistable cooling function $t=1.99\times10^6\:GM/c^3$ and at an arbitrary time when the disk is truncated ($t=5.96\times10^5 \:GM/c^3$).  In the bottom row, the time-averaged density and temperature maps are shown of the portion of the simulation over which we the used in our analysis ($t>1.99\times10^6\:GM/c^3$).  Overlaid is the average geometric scale height ($h_g/r$, black line) and the temperature scale height ($h_T/r$, white dashed line). 
\label{fig-den_ave}}
\end{figure*}

The gas in the inner region of the disk responds to the change in cooling by inflating since it is beyond $T_{T}$ and has an excess of thermal energy which it cannot rapidly dissipate on the inefficient cooling branch.  Additionally, viscous stresses continue to do work on the gas, further increasing the temperature.  A thin, dense stream of material plunges into the inner region between $t=2.01\times10^6\:GM/c^3$ and $t=2.04\times10^6\:GM/c^3$.  Since this gas is near $T_{T}$, it ablates away, ultimately leaving a truncated disk.  The final remnants of the transient can be seen in a very thin stream of dense gas in the inner region of the disk at $t=2.01\times10^6\:GM/c^3$ and large, buoyant bubbles rising away from the transition radius.

The abrupt change in the cooling law is, obviously, unphysical and the heating event is the reaction of the gas to the sudden switch.  Following this, however, the disk quickly settles into a new, sustained configuration with a ``transition zone" from $r=80-100\:r_g$.  In this transition zone, the region is cyclically filled with material through accretion, but then evacuated as the gas heats to $T_{T}$.  \emph{Unless explicitly stated, the analysis we detail is conducted following the implementation of the bistable cooling function.}

Figure \ref{fig-den_ave} shows two representative snapshots of the gas density and scale height temperature during initialization and during the analysis.  Before the change in cooling function, the disk has a uniform scale height ratio and very little hot gas is present in the simulation.  In the density and temperature maps of the truncated disk, wispy regions of cold, dense gas are seen surrounding diffuse, hot bubbles.  The cold features are very dynamic and short lived, as they are close to the transition temperature and can spontaneously transition to the hot phase.

Figure \ref{fig-den_ave} also shows the time averaged density of the disk with two measures of the characteristic disk scale height overlaid.  The first measure is that of the geometric scale height ratio, \begin{equation}
\frac{h_{g}(r)}{r} = \Bigg\langle \sqrt{\frac{\int (\theta(r) - \bar{\theta}(r))^2 \rho d\Omega}{\int \rho d\Omega}} \Bigg \rangle,
\end{equation} where $d\Omega=\sin \theta d\theta d\phi$ is the solid angle element in spherical coordinates and , \begin{equation}
\bar{\theta}(r) = \frac{\int \theta(r) \rho d\Omega}{\int \rho d\Omega}
\end{equation} is the average polar angle of the gas.  The second measure is the density weighted, temperature scale height ratio,
\begin{equation}
\frac{h_{T}(r)}{r} = \Bigg\langle \frac{\int \sqrt{T r} \rho d\Omega}{\int \rho d\Omega} \Bigg\rangle,
\end{equation} where $T$ is the gas temperature, $T = P/\rho$.  The values of $h_{g}/r$ and $h_{T}/r$ were calculated for every data dump and time averaged, denoted by the angled brackets.  \emph{Throughout this paper we take gas in the ``hot phase" to be that with $T>T_T$ and gas in the ``cold phase" to have $T<T_T$.}

Spatially, the time-averaged gas density shows a distinct truncation at $r=100$ $r_{g}$, which also corresponds to location where the disk becomes thicker, as shown by $h_{g}/r$ and $h_{T}/r$.  In the inner region of the disk $\rho$ is lower than the disk body by approximately an order of magnitude.  The averaged scale height ratio of the inner disk is only twice as thick as the rest of the disk and does not reach the ``thick" scale height target of $h_h/r=0.4$ due to a significant amount of advective cooling.

\begin{figure}
\includegraphics[width=0.5\textwidth]{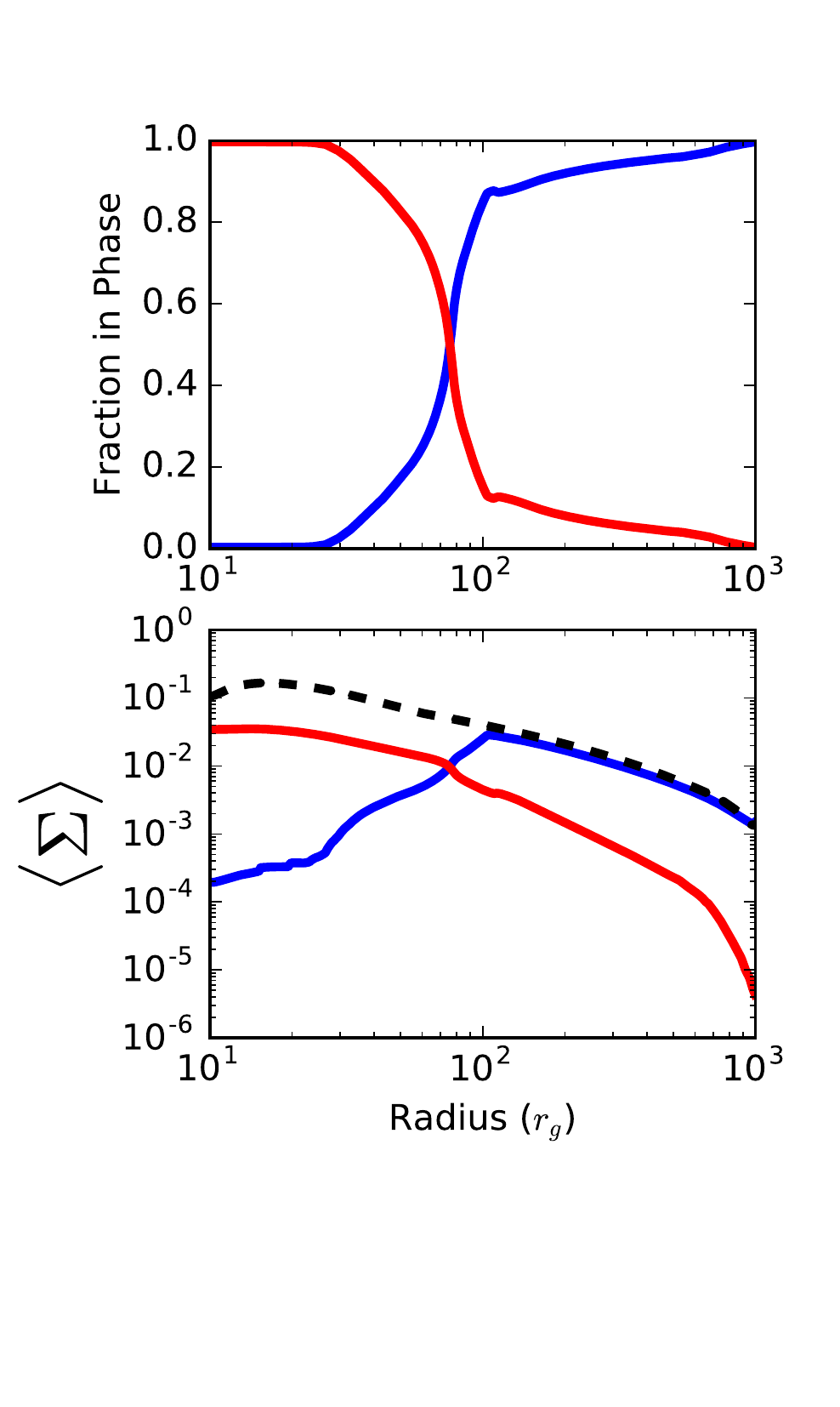}
\caption{Top panel- Radial profiles of the time-averaged $f_c$ (blue line) and $f_h$ (red line).  Bottom panel- Radial profiles of time-averaged $\Sigma_c$ (blue line) and $\Sigma_h$ (red line).  The surface density at the end of the initialization prior to the switch to the bistable cooling function is shown as the black dashed line.
\label{fig-phase_frac}}
\end{figure}

The signature of the thick-to-thin transition is also present in the thermal profile of the disk.  Figure \ref{fig-phase_frac} shows the time averaged radial profile of the fraction of gas in the hot and cold phases.  The fractions do vary in time, but, nevertheless, from this profile, we can see hot phase gas typically resides in the inner region of the disk and cold phase gas resides beyond the density transition.  It is also worth noting that the transition from a predominately hot to cold phase, i.e. where the fraction gas in the hot phase equals the fraction of gas in the cold phase, does not occur at $r=100\:r_{g}$ as we would expect from the cooling function, but rather just inwards of this radius at $r=80\:r_{g}$.  In brief, this is due to delayed heating as the gas is close to, but not yet at, $T_{T}$ as it crosses $r=100\:r_{g}$.  At the hot/cold flow interface it moves inwards and then makes the transition within the expected transition radius.

The truncation is also present in the time averaged surface density, \begin{equation}
\Sigma=\frac{1}{2 \pi r}\int \rho r^2 d\Omega.
\end{equation}  The hot and cold gas phases as a function of radius is also shown in Figure \ref{fig-phase_frac}.  Compared to the surface density at the end of the initialization, the inner disk surface density is lower by a factor of five.  Further out, beyond the transition radius where the gas is predominantly in the cold phase, the overall surface density is very near the intialization level.  Thus, through the launching of the convective cells by the transitioning gas, there is essentially an evacuation of the inner disk relative to the untruncated disk.

\subsection{Disk Dynamics}
\label{sec-disk_dynamics}

The disk dynamics in a truncated disk are of interest since hot and cold accretion flows should behave quite differently, but the truncated disk model used to explain astrophysical sources expects the two flows to exist in close proximity.  In the canonical thin accretion disk, the disk is rotationally supported and thermal motions are a small component of the overall velocities, i.e. $h/r=c_s/v_K$ where $v_{K}$ is the Keplarian velocity \citep{1973A&A....24..337S}.

\begin{figure}
  \subfigure[$\langle v_{\phi} \rangle / v_{K}$]{\includegraphics[width=0.5\textwidth]{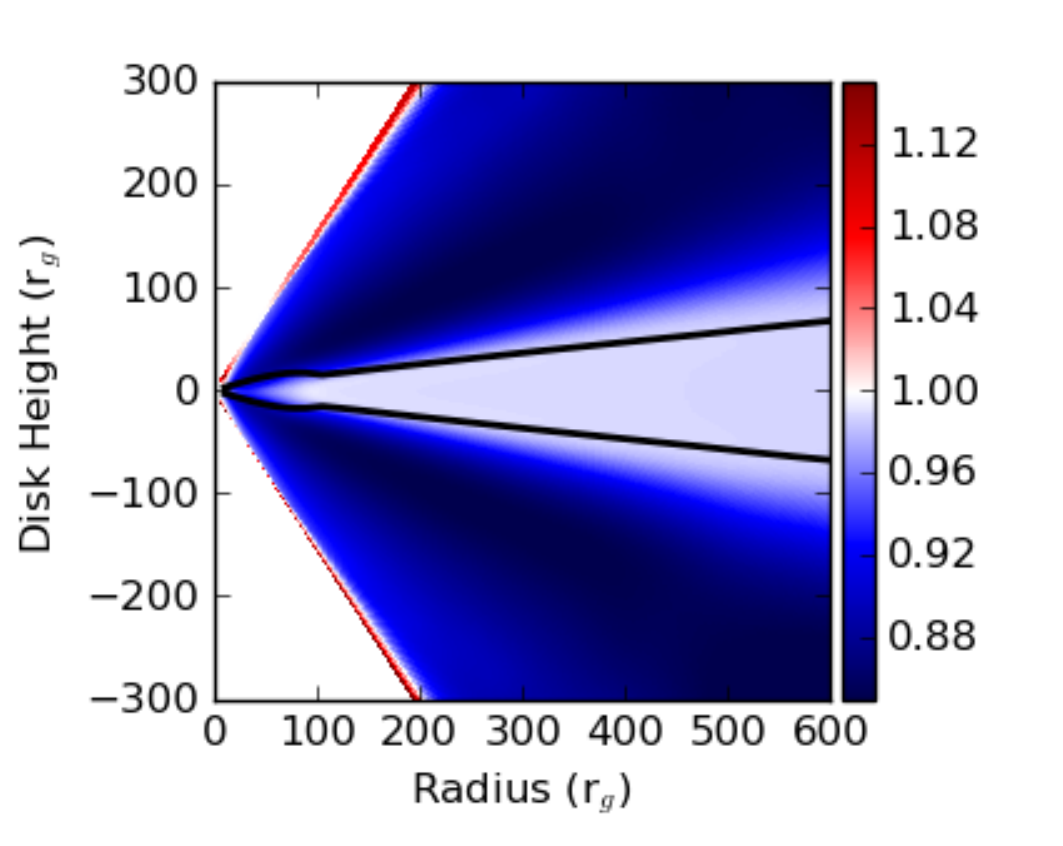}}
   \subfigure[$\langle v_{r} \rangle / v_{K}$]{\includegraphics[width=0.5\textwidth]{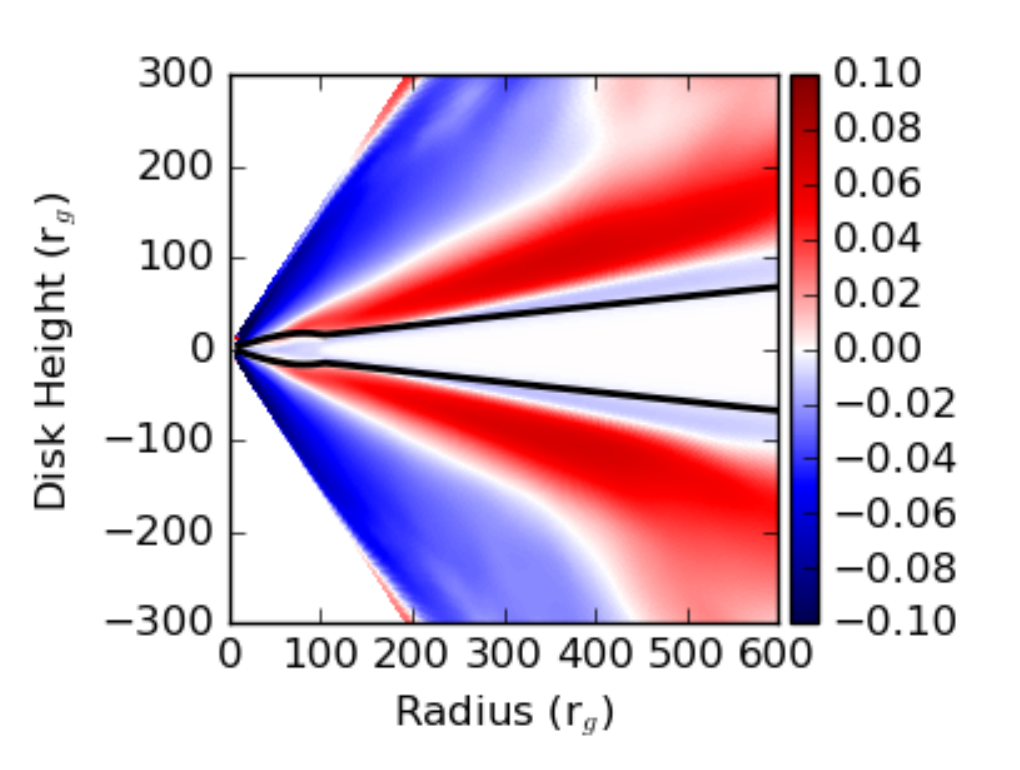}}
\caption{Maps of $\langle v_\phi \rangle / v_k$ and $\langle v_r\rangle / v_k$.  For reference, $h_g / r$ is overlaid.  The $\langle v_\phi \rangle / v_k$ map shows the gas in the coronal regions of the disk and within the transition radius is sub-Keplerian.  The stratified radial velocity structure of the gas is seen in the $\langle v_r\rangle / v_k$ map.  Within the cold disk body, the gas has very little radial motion.  The inflowing gas is seen as a thin boundary layer at the geometric scale height.  Large convective cells are found the coronal regions of the disk, which are seen here as alternating positive velocities at lower altitudes and negative velocities at higher altitudes.
\label{fig-vel_ratios}}
\end{figure}

In Figure \ref{fig-vel_ratios} the time averaged ratios of $\langle v_{\phi} \rangle / v_{K}$ and $\langle v_{r} \rangle / v_{K}$ are shown as a function of position in the simulation domain.  In the normalized simulation units (i.e. $G=M=c=1$), the Keplerian velocity is: \begin{equation}
v_{K} = \frac{1}{\sqrt{r}}.
\end{equation}

The disk establishes four dynamically distinct regions.  First is the outer disk body where the gas follows Keplerian orbits with very small radial velocity, as in a standard thin accretion disk.  Second, are the coronal regions, where the gas is sub-Keplerian with an average azimuthal velocity of $\approx15\%$ below $v_{K}$.  For comparison, previous hydrodynamic studies of hydrodynamic CDAF-like flows \citep[e.g.][]{1999MNRAS.310.1002S}, typically find that the azimuthal velocities are nearly-Keplerian and greater than we find here.  Additionally, as can be seen in the plot of $\langle v_{r} \rangle / v_{K}$, there is a substantial outflow with $v_r \approx 10\%$ of the local $v_K$ immediately above and below the disk, and a comparable fallback velocity at higher altitude.  This is due to a single convective cell that occupies each of the coronal regions.  Third is the inner disk body found within $r=100 \: r_g$ and one scale height above and below the disk.  In Figure \ref{fig-den_ave} we saw that this region was less dense, but contained a fair amount of cold gas.  In Figure \ref{fig-vel_ratios} we can see that this cooler region is modestly sub-Keplerian and inflowing at $1\%$ of $v_K$.  Finally, there is a thin layer of gas enveloping the disk at one scale height where the $\langle v_{r} \rangle / v_{K}$ map reveals that there is a prominent inflow directly on the disk body.  It is in this region that the bulk of the mass accretion from the outer disk occurs.

The launching mechanism is interesting, especially considering this is a simple hydrodynamic simulation so a magnetocentrifugally driven wind \citep{1982MNRAS.199..883B} or a radiation driven wind \citep{2000ApJ...543..686P, 2004ApJ...616..688P} cannot form.  Here, we find the outflow in the simulation is launched by pockets of hot gas that spontaneously undergo a transition from the cold phase to the hot phase in the cool stream.  As the gas expands, it becomes less dense and, thus, buoyantly unstable according to the the Solberg-H{\o}iland criterion describing the stability of a rotating fluid.  When the gas is launched, there is already an angular momentum deficit, so it has little rotational support at a large radius.  Unlike a pure CDAF, the gas accretes when it falls back because, while the cooling of the hot gas is inefficient, it is not zero and Bernoulli parameter is not conserved.

\begin{figure}
\includegraphics[width=0.5\textwidth]{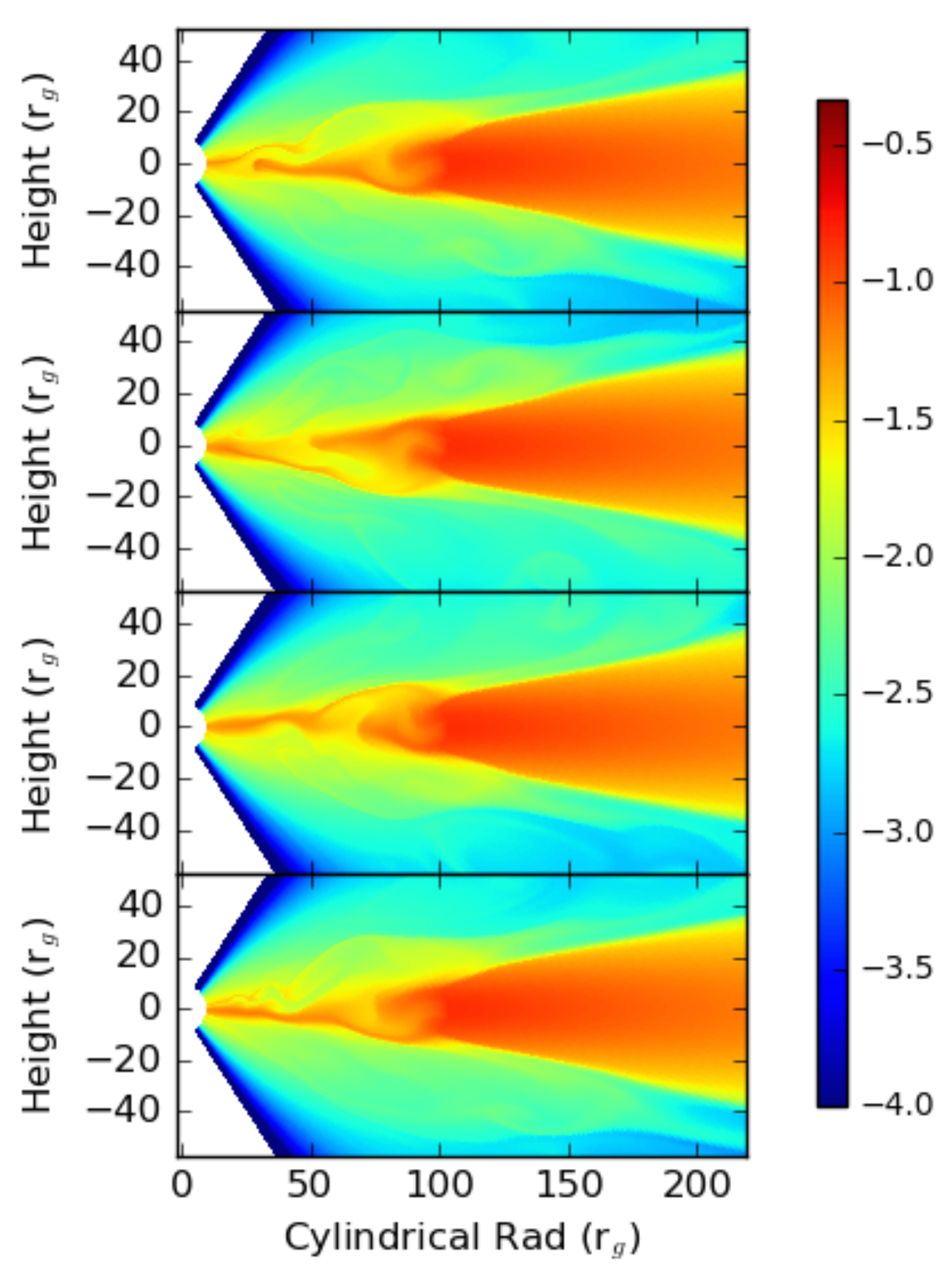}
\caption{Launching of outflow as pockets of hot gas spontaneously transition to the hot phase.  This sequence of images shows the gas density in the inner region at $t=2.48\times10^6\:GM/c^3$ (12,500 inner orbits), $2.50\times10^6\:GM/c^3$ (12,600 inner orbits), $2.52\times10^6\:GM/c^3$ (12,700 inner orbits), and $2.54\times10^6\:GM/c^3$ (12,800 inner orbits) and is representative of the general behavior observed in the simulation. 
\label{fig-bubble}}
\end{figure}

The launching processes is shown in a sequence of images in Figure \ref{fig-bubble}.  The cold, sub-Keplerian stream can be seen in the dense gas.  As the gas spontaneously transitions to the inefficient cooling branch, it heats and expands.  In doing so, these pockets of newly heated gas are surrounded by denser, cooler material so they buoyantly rise which disrupts the cold stream and disturbs the flow.  Once the gas reaches a low density region, at approximately one density scale height, it moves radially against the temperature-gravity gradient.

\begin{figure*}
\centering
\includegraphics[width=0.9\textwidth]{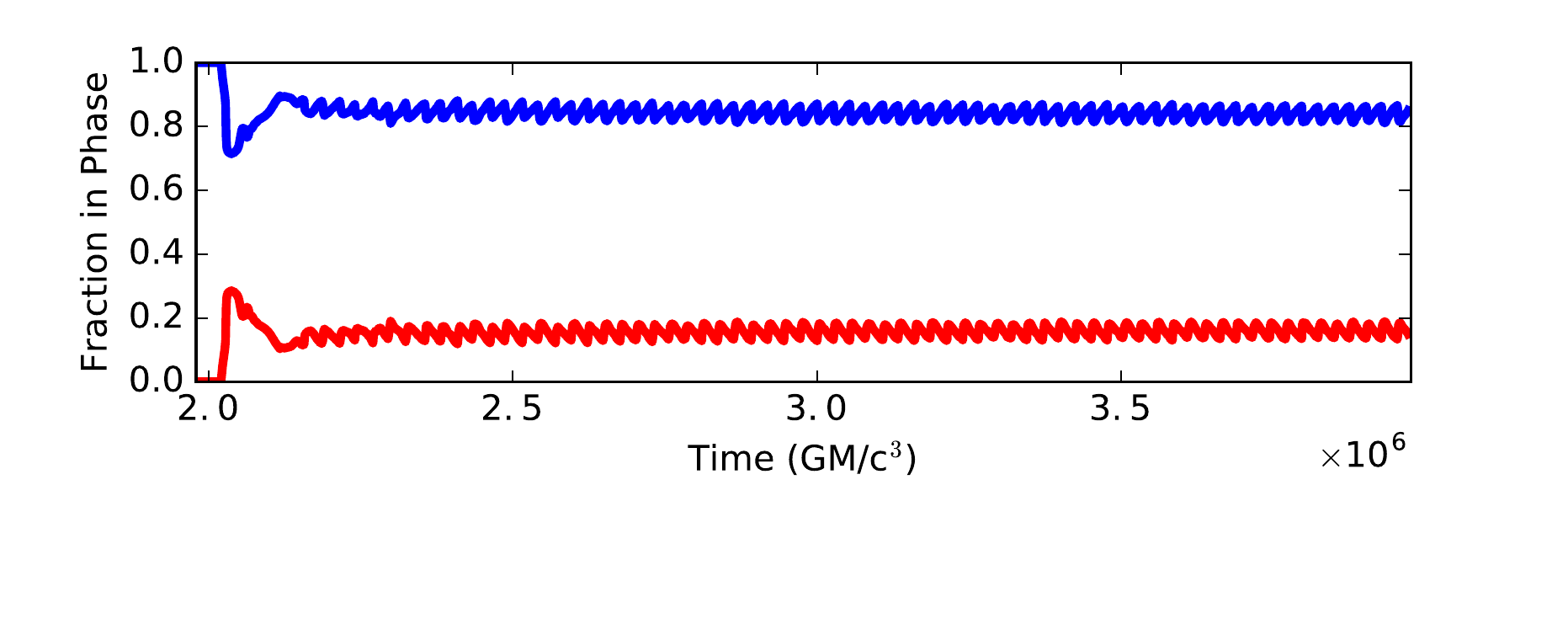}
\caption{Gas fractions ($f_c$ and $f_h$) at $r=100\:r_g$ following the switch to the bistable cooling function.  The initial heating event is seen as a spike in the hot gas fraction and subsequent decay at early times.  The gas fraction oscillations are found after this transient behavior dies away.
\label{fig-frac_osc}}
\end{figure*}

Figure \ref{fig-frac_osc} shows a time trace of the fraction of the gas in the hot phase, $f_h$, and the fraction of gas in the cold phase, $f_c$, at $r=100 \: r_g$.  The transient heating event is seen at early times in time traces of $f_h$ and $f_c$.  Once the disk relaxes into its truncated configuration, $f_h$ and $f_c$ oscillate as heating events quasi-periodically occur in the cold streaming material and hot gas bubbles rise.  These fluctuations are relatively small, but rapid with the fluctuations on average taking $t=2.65\times10^4\:GM/c^3$, or four thermal times.  At $r=100\:r_g$ the average gas fractions in each phase are $f_c=0.84$ and $f_h=0.15$.  The oscillations in the gas fraction are asymmetric and have a standard deviation of $0.03$.  As the oscillations occur, $f_h$ rises rapidly and then experiences a slower decay.  Consequently, $f_c$ decreases sharply and gradually increases before the cycle repeats.

\subsection{Mass Accretion and Angular Momentum Transport}
\label{sec-ang_mom}

\begin{figure*}
\centering
\includegraphics[width=0.9\textwidth]{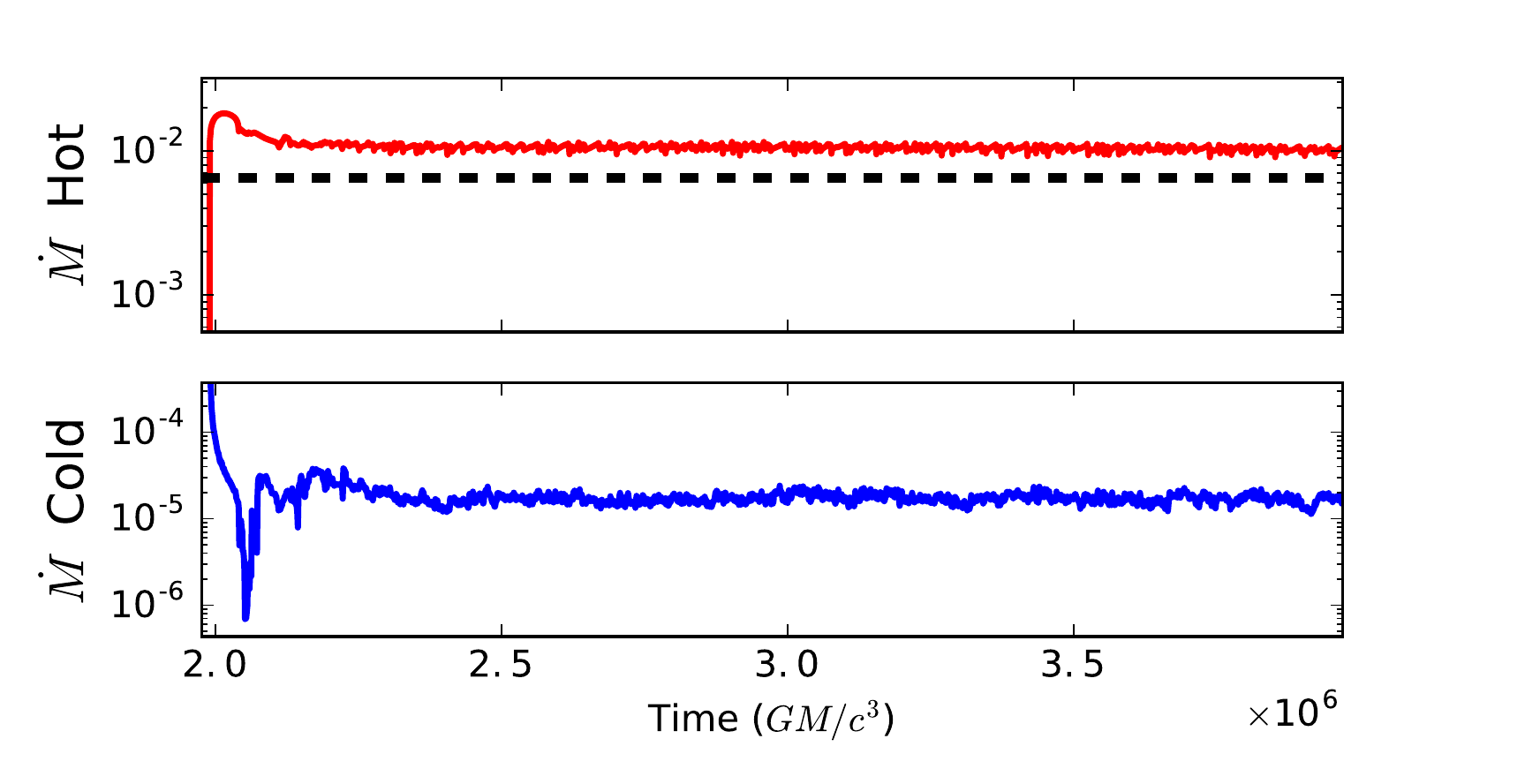}
\caption{$\dot{M}_h$ (top panel) and $\dot{M}_c$ (bottom panel).
\label{fig-frac_mdot}}
\end{figure*}

Previous numerical studies of convective-like, hydrodynamic accretion flows have found that very little gas actually accretes onto the black hole \citep{1999MNRAS.303..309I, 1999MNRAS.310.1002S, 2000ApJS..130..463I, 2000ApJ...539..798N}.  Here, we report conflicting results, as we find that the hot gas does indeed accrete in our simulation.  One possibility for this divergence from previous studies of hot flows is that the convective cells in our simulation are continually fed at the truncation of the thin disk so the gas originates from deeper in the the gravitational potential, rather than being a hot flow throughout the entire domain and overall being less gravitationally bound.

In our 2D, axisymmetric disk, the instantaneous mass accretion rate in the simulation is given by,\begin{equation}
\dot{M}(R) = \int -\rho v_{r} r^2 d\Omega.
\end{equation}  Using our definition of the gas phases from Section \ref{sec-trunc}, we show the time trace of the accretion rate for the hot ($\dot{M}_H$) and cold phase ($\dot{M}_C$) gas at the inner boundary of our simulation in Figure \ref{fig-frac_mdot}.  The pre-analysis disk mass accretion rate of cold gas, $\langle \dot{M}_{pre} \rangle = 6.65\times10^{-3}$, is also shown for reference. 

We find, somewhat surprisingly, the mass flux across the inner boundary is slightly higher in the truncated disk than in the initialization phase.  Additionally, there is a striking disparity between the $\dot{M}_h$ and $\dot{M}_c$.  Overwhelmingly, by approximately three orders of magnitude, hot gas is accreted rather than cold gas.  Initially, the overall mass accretion rate is higher during the transient phase of the heating event, but it settles into a sustained accretion rate of $\langle \dot{M} \rangle = 1.0\times10^{-2}$ for the remaining length of the simulation.

\begin{figure}
\includegraphics[width=0.5\textwidth]{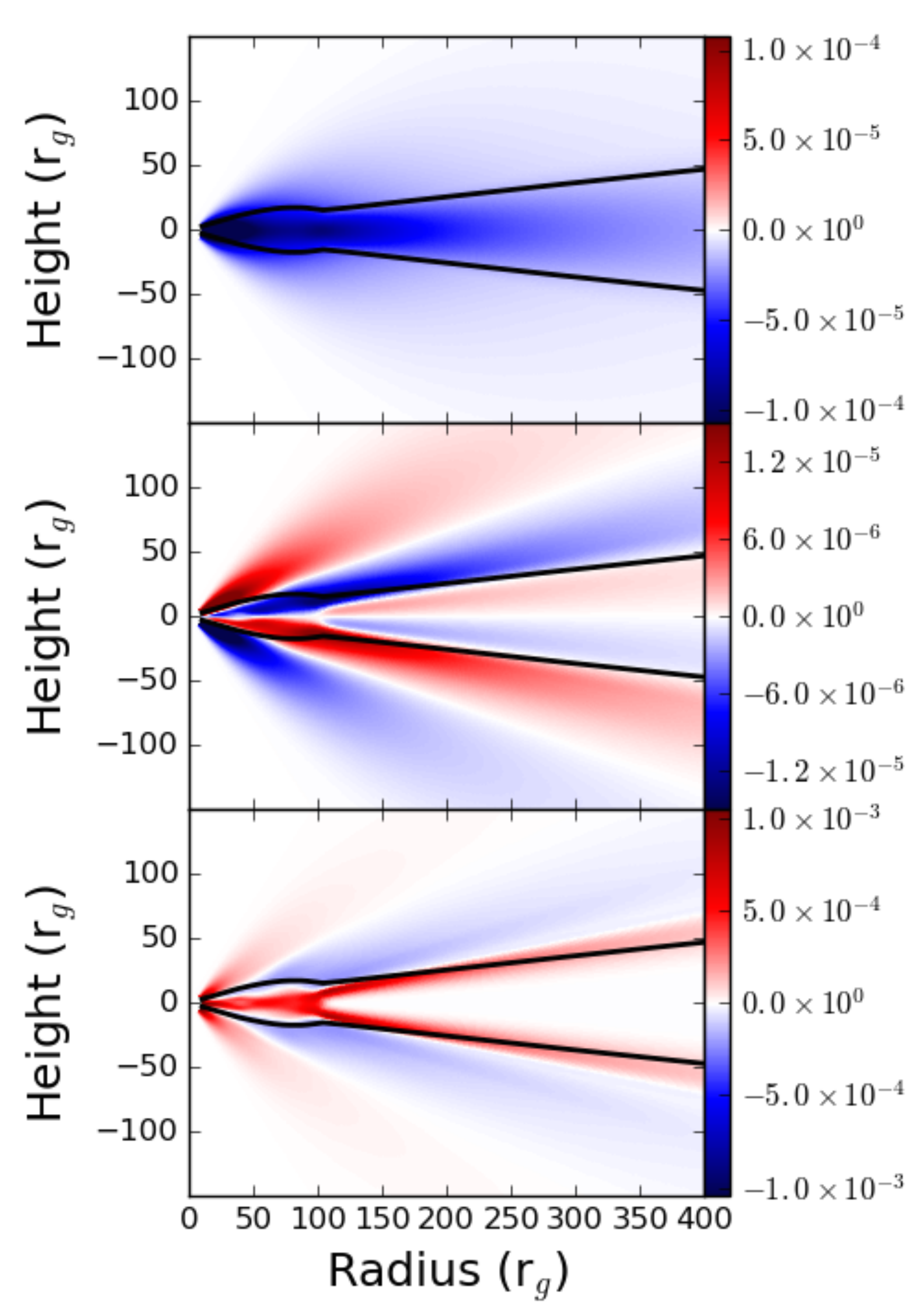}
\caption{Maps of $r-\phi$ viscous stress (top panel), $\theta-\phi$ viscous stress (middle panel), and advection of angular momentum (bottom panel).  Overlaid is the average geometric scale height (black solid line).
\label{fig-stress_comps}}
\end{figure}

Since an appreciable amount of gas is accreted, there must be efficient angular momentum transport in the simulation.  In our disk, the angular momentum can be transported three ways: through advection, stresses arising from the radial shear of the disk, and stresses from shear in the atmosphere of the disk as it rotates within the sub-Keplerian convective outflow.  The conservation of angular momentum equation takes the following form, \begin{eqnarray}\label{eqn-ang_mom}
\frac{\partial}{\partial t} \bigg( \rho r v_\phi \bigg) = \nabla \cdot \bigg(\eta r^2 \frac{\partial}{\partial r}\bigg( \frac{v_\phi}{r \sin\theta}\bigg)\uvec{e}_r \\ + \eta \sin\theta \frac{\partial}{\partial \theta} \bigg(\frac{v_{\phi}}{\sin \theta}\bigg)\uvec{e}_\theta -\rho r v_\phi \vvel \bigg),\nonumber
\end{eqnarray} where the three components appearing in the divergence term are, respectively, the advection term, the $r-\phi$ viscous stress term, and the $\theta-\phi$ stress term.  The time-averaged maps of these terms shown in Figure \ref{fig-stress_comps}.

\begin{figure}
\includegraphics[width=0.5\textwidth]{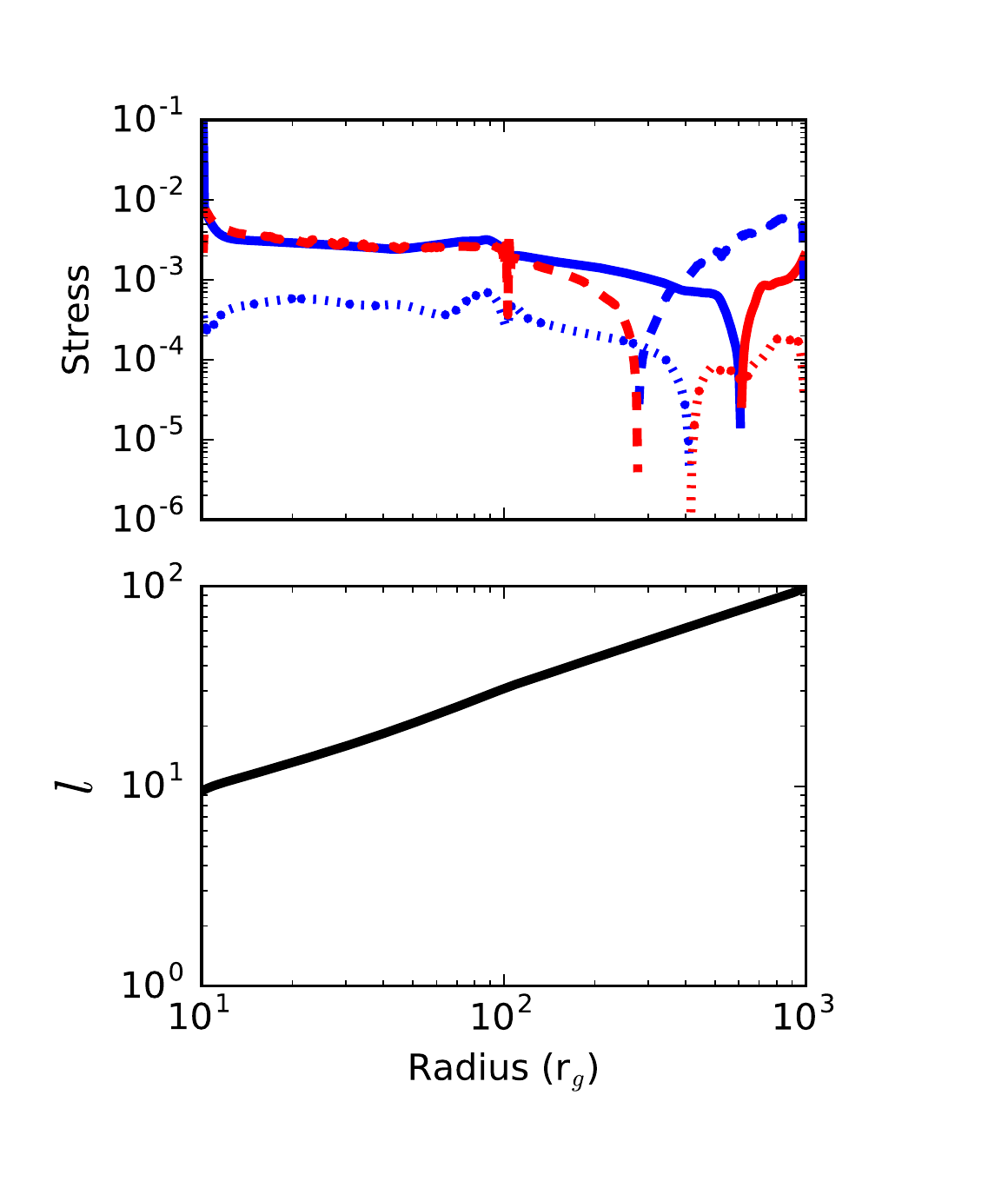}
\caption{Top panel- Radial profiles of angular momentum advection (dashed line), $r-\phi$ viscous stress term (solid line), and $\theta-\phi$ term (dotted line) integrated through a shell at a given radius.  Blue indicates a negative quantity and red indicates a positive quantity.  Bottom Panel- Radial profile of time averaged specific angular momentum.
\label{fig-stress_prof}}
\end{figure}

The importance of the transport terms at different locations in the disk is quite clear.  The $r-\phi$ viscous stress is negative everywhere and largest within the disk body, closer in towards the inner domain.  This structure is expected since we consider an $\alpha$-type viscosity where the viscous stress is proportional to the gas pressure and $\phi$-shear and the pressure is largest in the inner regions of the disk.  Above and below the disk, at approximately one scale height, there are localized layers of  $\theta-\phi$ stress where the Keplerian disk interacts with the sub-Keplerian outflow.  Also, within the very inner regions, the sign of the stress in each respective hemisphere reverses where the convective cell closes, reflecting the opposite signs of the vertical $v_\phi / \sin (\theta)$ gradient.  The largest $\theta-\phi$ stresses occur on the upper and lower edges of the truncated disk where the cold, dense disk is buffeted by the outflowing, hot, buoyant gas.  The stress gradually decreases at larger radii as the strength of the interaction between the disk and outflow decreases. 

\begin{figure*}
\includegraphics[width=\textwidth]{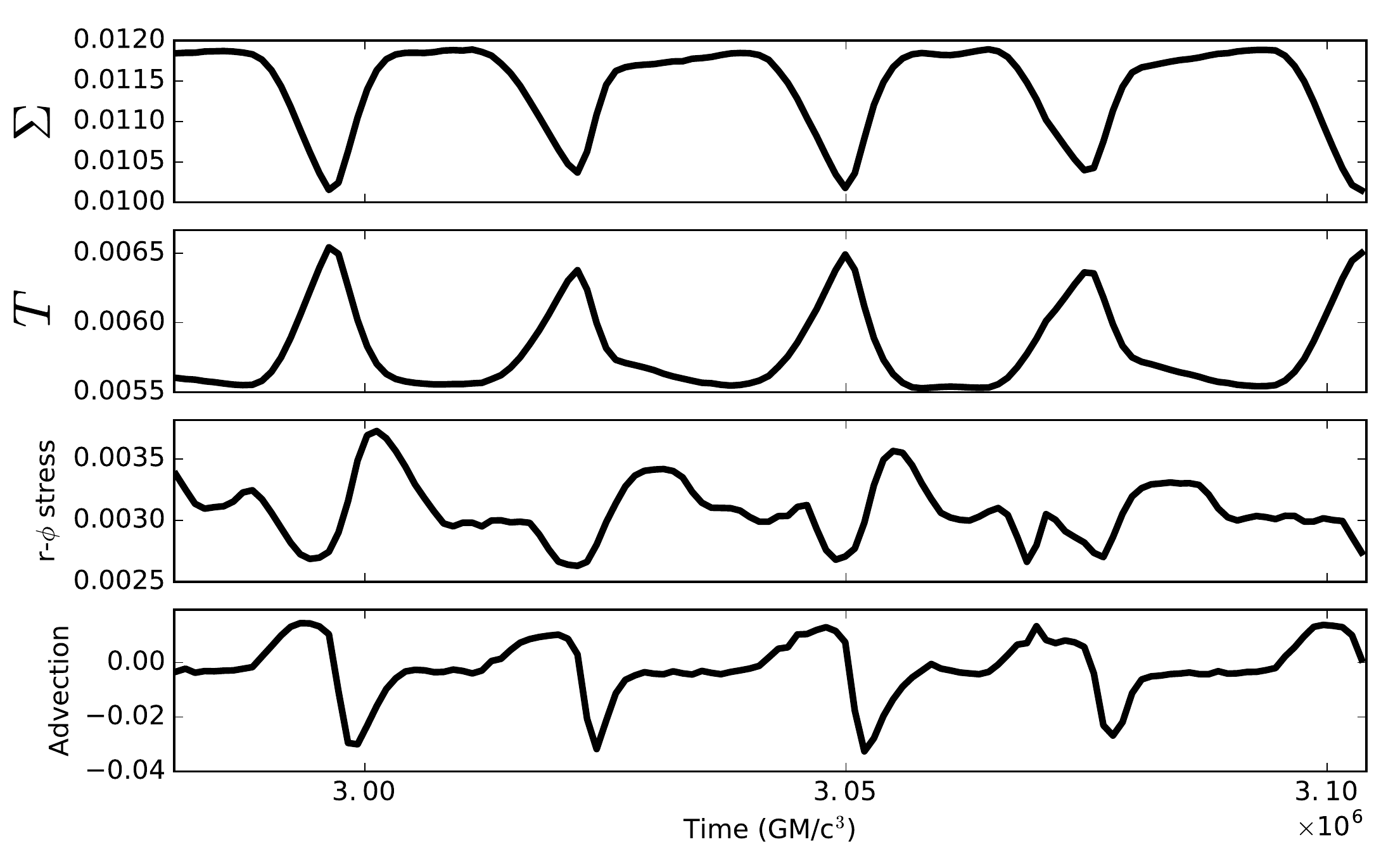}
\caption{Time variability of $\Sigma$ (top panel), $T$ (second panel), $r-\phi$ stress (third panel), and advection (bottom panel) at $r=80\:r_g$.
\label{fig-ang_mom_time}}
\end{figure*}

The time averaged map of the angular momentum advection terms is also stratified, reflecting the structure seen in Figure \ref{fig-vel_ratios} of the radial velocities.  It is worth noting that within the disk body meridional circulation establishes a very small, but positive, net radial velocity at the disk midplane.  This behavior was originally predicted by \citet{1982A&A...114...77K} and \citet{1984SvA....28...50U} and has since been reproduced numerically for viscous accretion disks \citep{1992ApJ...397..600K, 1994ApJ...423..736R}.  Moving away from the disk midplane to higher latitude, the circulation reverses the flow and gas moves inwards.

The radial profiles of the time averaged angular momentum fluxes resulting from the terms in Equation \ref{eqn-ang_mom} are shown in Figure \ref{fig-stress_prof}.  On average, the viscous $r-\phi$ stress is balanced by advection, at least within the region that has established a quasi-steady state and viscous equilibrium (out to $\approx 500\:r_g$).  In a purely convective flow, angular momentum is advected inwards, rather than transported outward \citep{1992ApJ...388..438R, 1996ApJ...464..364S}.  However, in this model, the viscous stresses continually act on the gas, allowing for the loss of angular momentum.  Consistent with the removal of angular momentum by viscous torques, the specific angular momentum profile, also shown in Figure \ref{fig-stress_prof}, is Keplerian, $l\propto r^{1/2}$, when averaged over the duration of the simulation.

The angular momentum transport at the hot-cold phase interface is complex and depends on the time variability of $\Sigma$, $T$, the $r-\phi$ stress component, and the advective stress component.  Figure \ref{fig-ang_mom_time} shows these quantities at $r=80 r_{g}$, in the heart of the transition zone, over several of the bubble launching cycles.  In Figure \ref{fig-ang_mom_time}, $\Sigma$, $T$, the $r-\phi$ stress, and advective stress have been averaged over one thin-disk scale height ($h=\pi/2\pm0.1$). 

At the truncation, a cyclical building and expulsion of material occurs with a period of $t\approx2.65\times10^4\:GM/c^3$.  In this cycle, peaks in $T$ occur when the density is lowest and vice versa.  The angular momentum terms vary with $\Sigma$ and $T$, but there is a slight delay, $\Delta t \approx2.05\times10^3\:GM/c^{3}$, in the maxima and minima.  For the majority of the cycle, approximately $80\%$, the surface density increases.  Eventually, a critical threshold is reached when the heating rate exceeds the cooling rate, causing the temperature to increase and the cooling to become inefficient as pockets of gas reach $T_{c}$.  This causes a runaway heating event, seen in the rapid increase of $T$, and the outflow to be launched which can be seen as turn over in the advection term to positive values.

\begin{figure}
\includegraphics[width=0.5\textwidth]{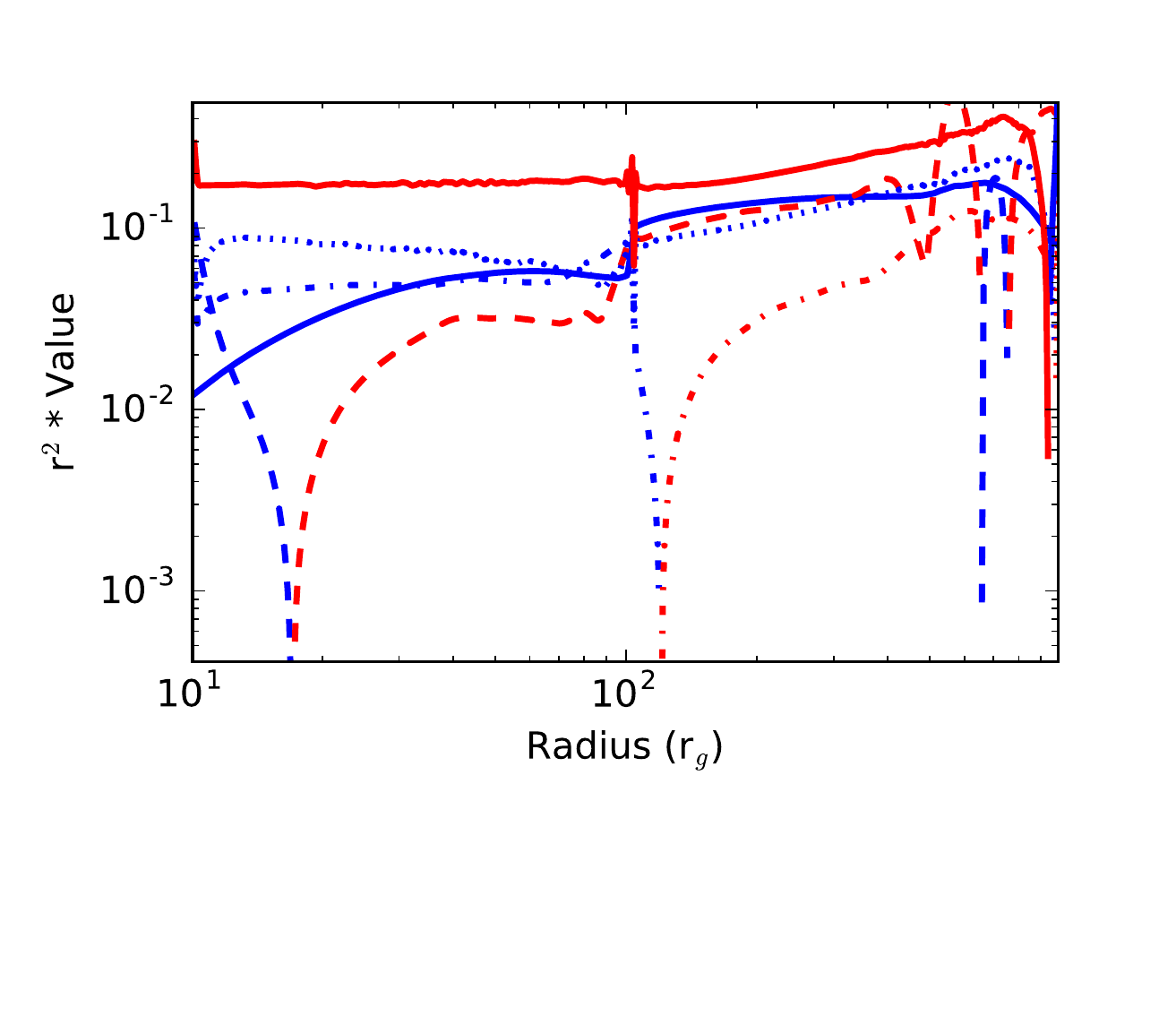}
\caption{Radial profiles of terms in energy equation.  Blue indicates negative values, red indicates positive values.  The combined viscous dissipation ($\mathcal{W}$, $\theta-\phi$ and $r-\phi$ terms) are shown by the dashed line, the advection of kinetic energy ($\mathcal{A_K}$) is shown by the dotted line, and the advection of thermal energy ($\mathcal{A_{T}}$) is shown by the dot-dashed line.  Since the cooling term ($\mathcal{C}$) is positive everywhere in our notation and the change in gravitational potential energy ($\mathcal{G}$) is negative everywhere, both are shown as solid lines.
\label{fig-energy_profile}}
\end{figure}

\begin{figure*}
\centering
\includegraphics[width=\textwidth]{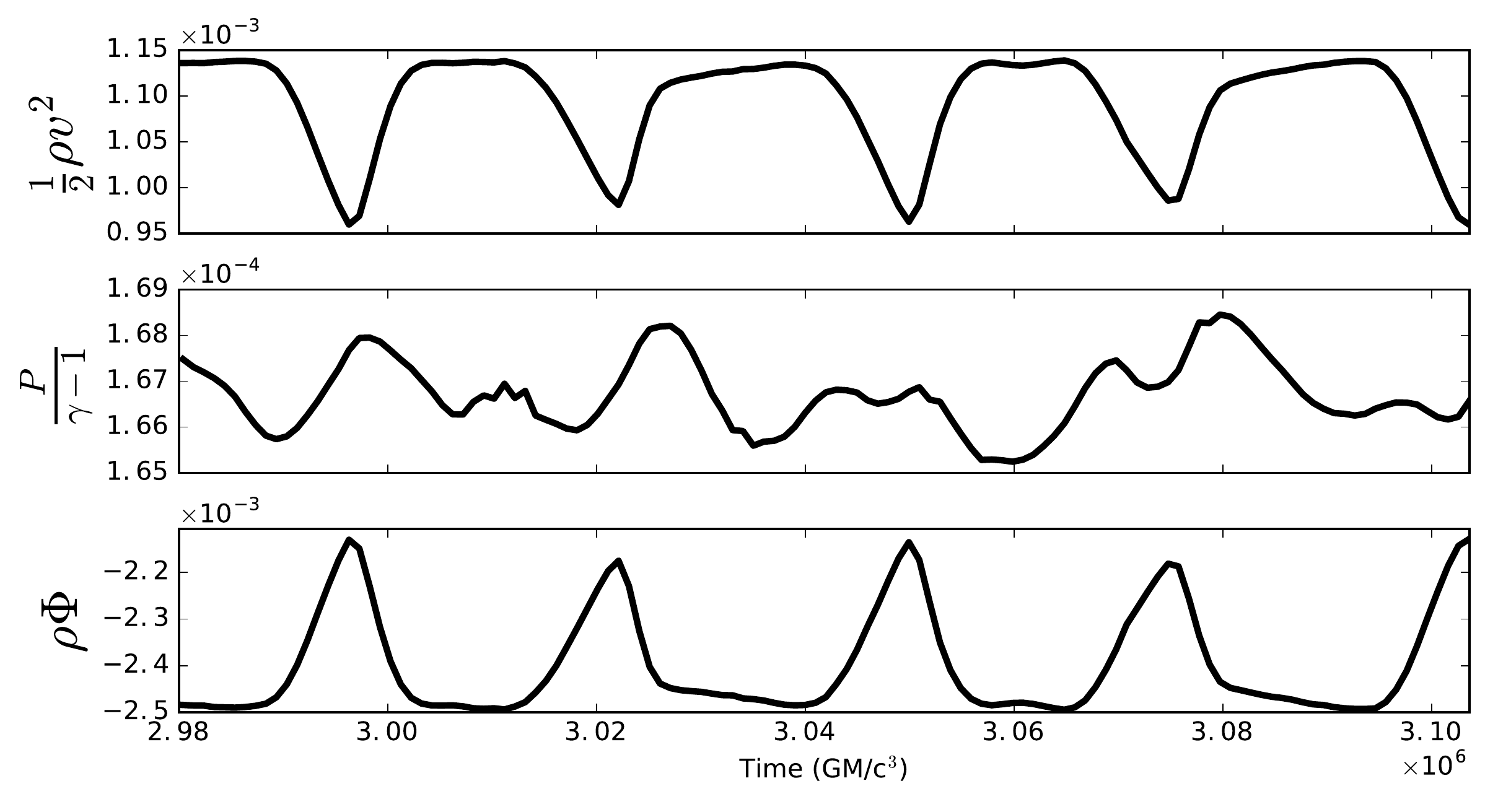}
\caption{Time variability of $\mathcal{E}_K$ (top panel), $\mathcal{E}_I$ (middle panel), and $\mathcal{E}_G$ (bottom panel) at $r=80\:r_g$.
\label{fig-energy_time}}
\end{figure*}

A key feature throughout this cycle is that the gas pressure remains relatively constant as the gas expands to maintain a pressure equilibrium, a facet which is discussed to greater depth in Section \ref{sec-energetics}.  This, in turn, means that the specific angular momentum of the lower density, transitioning gas at the truncation is more rapidly lost to the neighboring dense gas of the disk body through the $r-\phi$ viscous stress since the amplitude of the viscosity is proportional to the pressure.  Additionally, as seen in Figure \ref{fig-bubble}, the hot transitioning gas is in contact with the dense gas of the cooler stream which cannot lose angular momentum as well.  As the specific angular momenta of these two regions changes due to the differences in the viscous efficiency, a vertical shear is introduced.  This shear allows the $\theta-\phi$ viscous stress to transfer angular momentum from the dense gas in the stream to the diffuse, outflowing bubble where it is carried away.  This changes the shear rate of the dense gas, resulting in a spike in the $r-\phi$ stress component.  This rapid exchange of angular momentum drives the angular momentum transport in the stream which then flows to smaller radii, as evidenced by the sharp negative dip in the advective angular momentum transport term corresponding to the peak in the $r-\phi$ stress term.

After this involved transitory event, the disk refills and the cycle repeats.

\subsection{Disk Energetics}
\label{sec-energetics}

\subsubsection{Energy Terms}
In our hydrodynamic model, the conservation of energy equation is, \begin{equation}
\frac{\partial \mathcal{E}}{\partial t}  + \mathcal{C} + \mathcal{A_{K}} + \mathcal{A_{T}} + \mathcal{G} - \mathcal{W} = 0,
\end{equation} where $\mathcal{E}$ is the total disk energy, $\mathcal{C}$ is the ``radiative" cooling done by the cooling function, $\mathcal{A_{K}}$ is the advection of kinetic energy, $\mathcal{A_{T}}$ is the advection of thermal energy, $\mathcal{G}$ is the advection of gravitational potential energy, and $\mathcal{W}$ is the work done by viscous torques.  We will consider these terms in a spherical shell integrated form, such that they are  \begin{eqnarray}
\frac{\partial \mathcal{E}}{\partial t} = \frac{\partial}{\partial t} \bigg [ \int d\Omega \bigg(\frac{1}{2} \rho v^2 + \frac{P}{\gamma-1} + \rho \Phi \bigg)\bigg], \\
\mathcal{C} = \int d\Omega \Lambda, \\
\mathcal{A_{K}} = \frac{1}{r^2} \frac{\partial}{\partial r} \bigg[ r^2 \int d\Omega \bigg( \frac{1}{2} \rho v^2 \bigg)v_{r} \bigg], \\
\mathcal{A_{T}} = \frac{1}{r^2} \frac{\partial}{\partial r} \bigg[ r^2 \int d\Omega \bigg( \frac{P}{\gamma-1}\bigg)v_{r} \bigg], \\
\mathcal{G} = \frac{1}{r^2} \frac{\partial}{\partial r} \bigg[ r^2 \int d\Omega \bigg( \rho \Phi \bigg)v_{r} \bigg], \\
\mathcal{W} = \frac{1}{r^2} \frac{\partial}{\partial r} \bigg\{ r^2 \int d\Omega \eta v_{\phi} \bigg[ r \frac{\partial}{\partial r} \bigg(\frac{v_{\phi}}{\sin \theta} \bigg)
\\+ \sin\theta\frac{\partial}{\partial \theta}\bigg(\frac{v_{\phi}}{\sin\theta}\bigg)\bigg]\bigg\}. \nonumber
\nonumber
\end{eqnarray} 

The time averaged, radial profiles of these terms are shown in Figure \ref{fig-energy_profile}.  The presence of a truncated disk influences every energy term, with the exception of $\mathcal{G}$.  Given that there is sustained accretion and the average specific angular momentum profile is Keplerian, this is not necessarily surprising in the time average profile.  Essentially, the smooth, time averaged $\mathcal{G}$ indicates that, globally, material is accreting and there is no mass build-up anywhere in the disk.  Even though material may temporarily be caught in convective cells, it is eventually accreted so gravitational potential energy is liberated.

In the outer disk the $\mathcal{A_{T}}$ term, which is positive, is suppressed and accounts for very little of the overall energy transport of the gas.  This changes within $r<R_{T}$, however, as $\mathcal{A_{T}}$ and $\mathcal{A_{K}}$ become major contributors in the truncated disk region.  Advection plays a dominate role in the energy transport within the transition radius, overtaking the cooling.  The dynamical influence of the convective cells acts to advect energy across the inner boundary as material gets swept inwards.

There is a slight deficit in the $\mathcal{A_{K}}$ term and a significant deficit in the $\mathcal{W}$ term inside of $r<R_{t}$ because of the disk truncation.  The impact on the $\mathcal{A_{K}}$ term is twofold.  In the inner disk the density is lower and the flow is sub-Keplerian, thus overall advection of kinetic energy in this region is slightly less what it might otherwise be.  Regardless, the sign does not change as the bulk of the kinetic energy is swept inwards in the dense gas that streams inwards through the viscous interaction of the hot outflow.  In a general, time averaged sense, the energy dissipation of the $\mathcal{W}$ term is suppressed because the density is lower and the fluid motions are more chaotic in this region.  However, as seen in the variability of the viscous stress components, there are brief periods of enhanced viscous stress and, therefore, heating.  Nevertheless, the disruption of the flow by the hot gas bubbles makes it more difficult to establish the \emph{sustained} radial and vertical gradients in $v_\phi$ that generate the viscous stresses and heating in the outer disk, thereby diminishing the contribution of the $\mathcal{W}$ term in the inner region of the disk.

The transition has the strongest affect on the $\mathcal{A_T}$ term, as its sign actually flips at the disk transition.  Within the inner disk, material is generally swept inwards, either in the streaming material or through the return of the convective cells.  The sign of $\mathcal{A_T}$ flips at the transition, however, because the majority of the thermal energy resides in the buoyant, outflowing gas.  The apparent suppression of this term is somewhat artificial because it accounts for two competing components; the net inflow of thermal energy through the accretion in the dense, thin disk and the outflow of the diffuse, hot gas.  

Further insight into the periodic launching of the buoyant bubbles can be gained from examining the individual components of $\mathcal{E}$, i.e. the kinetic energy ($\mathcal{E}_K=1/2 \rho \vvel$), the internal energy ($\mathcal{E}_I=P/(\gamma-1)$), and the gravitational potential energy ($\mathcal{E}_G=\rho \Phi$) of the gas.  The variability of these energy components at $r=80\:r_g$ is shown in Figure \ref{fig-energy_time}.  Similar to Figure \ref{fig-ang_mom_time}, these terms have been averaged over $h=\pi/2\pm0.1$.

The cyclical behavior observed in $\Sigma$, $T$, the $r-\phi$ stress, and the angular momentum advection is also seen in the gas energy.  Unlike the involved angular momentum transport interaction observed in this region, the temporal variability of the gas energy terms is simply due to the removal of gas in the buoyant bubbles, which can be seen in the corresponding dips in $\Sigma$ in Figure \ref{fig-ang_mom_time} and in the variability of $\mathcal{E}_G$ since $\Phi$ is a constant at $r=80\:r_g$.  The removal of the gas is tied to an increase in the gas energy as it heats (i.e. increased $\mathcal{E}_I$).  Throughout the launching of the hot bubbles, the internal energy and gas pressure have very little variability ($\approx2\%$).  When gas transitions in the flow, the bubble expands adiabatically as it buoyantly rises vertically along the density gradient.

\subsubsection{Bound Outflow}
\label{sec-Bernoulli}

Throughout the simulation, the Bernoulli parameter, \begin{equation}
Be = \frac{1}{2} \vvel^2 + \Phi + W,
\end{equation} where $W$ is the specific enthalpy \begin{equation}
W = \frac{\gamma c_s^2}{\gamma - 1}
\end{equation} of the gas, is negative.  When $Be$ is negative, the kinetic and thermal energy of the gas is less than the gravitational potential energy, so the gas is bound and cannot escape.

\begin{figure}
\subfigure[Histogram of $Be$ of Outflowing Gas at $100\:r_g$]{\includegraphics[width=0.5\textwidth]{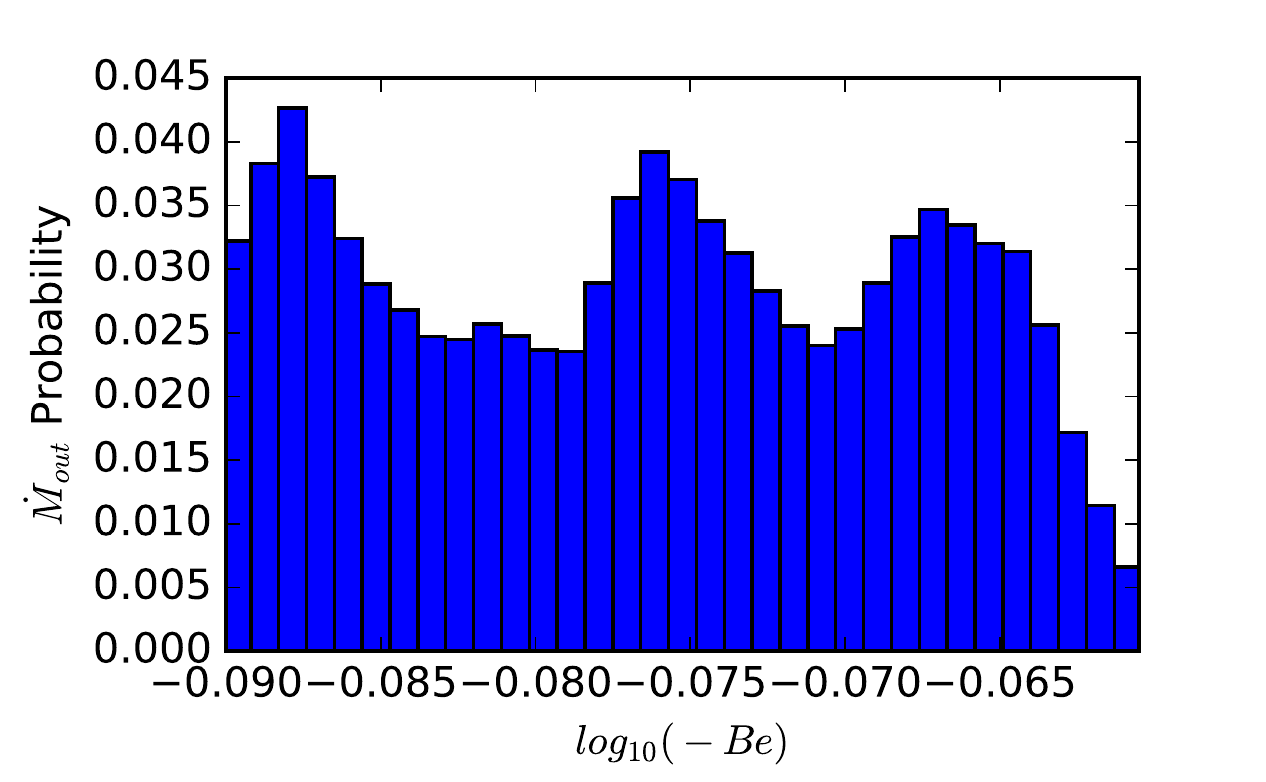}}
\subfigure[$Be$ vs. $\rho$ of Outflowing Gas at $100\:r_g$]{\includegraphics[width=0.5\textwidth]{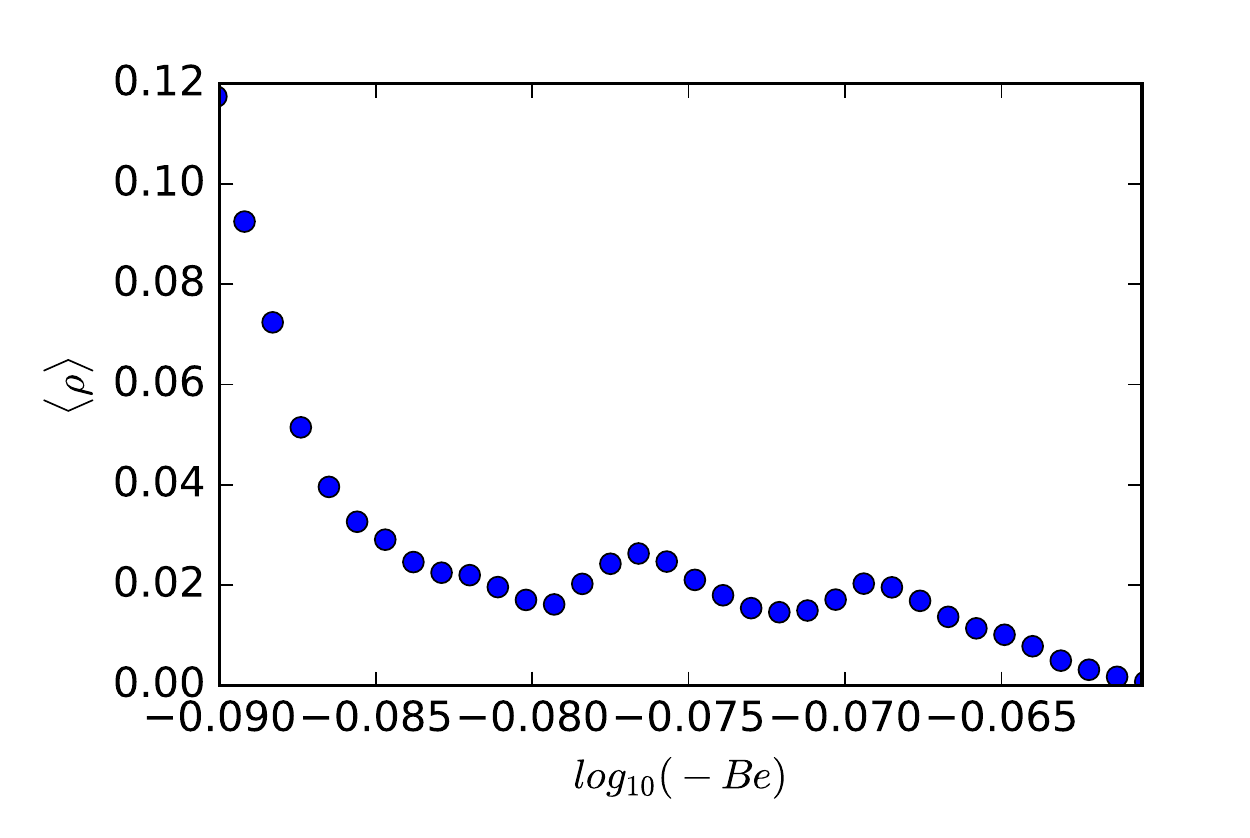}}
\subfigure[$Be$ vs. $v_r$ of Outflowing Gas at $100\:r_g$]{\includegraphics[width=0.5\textwidth]{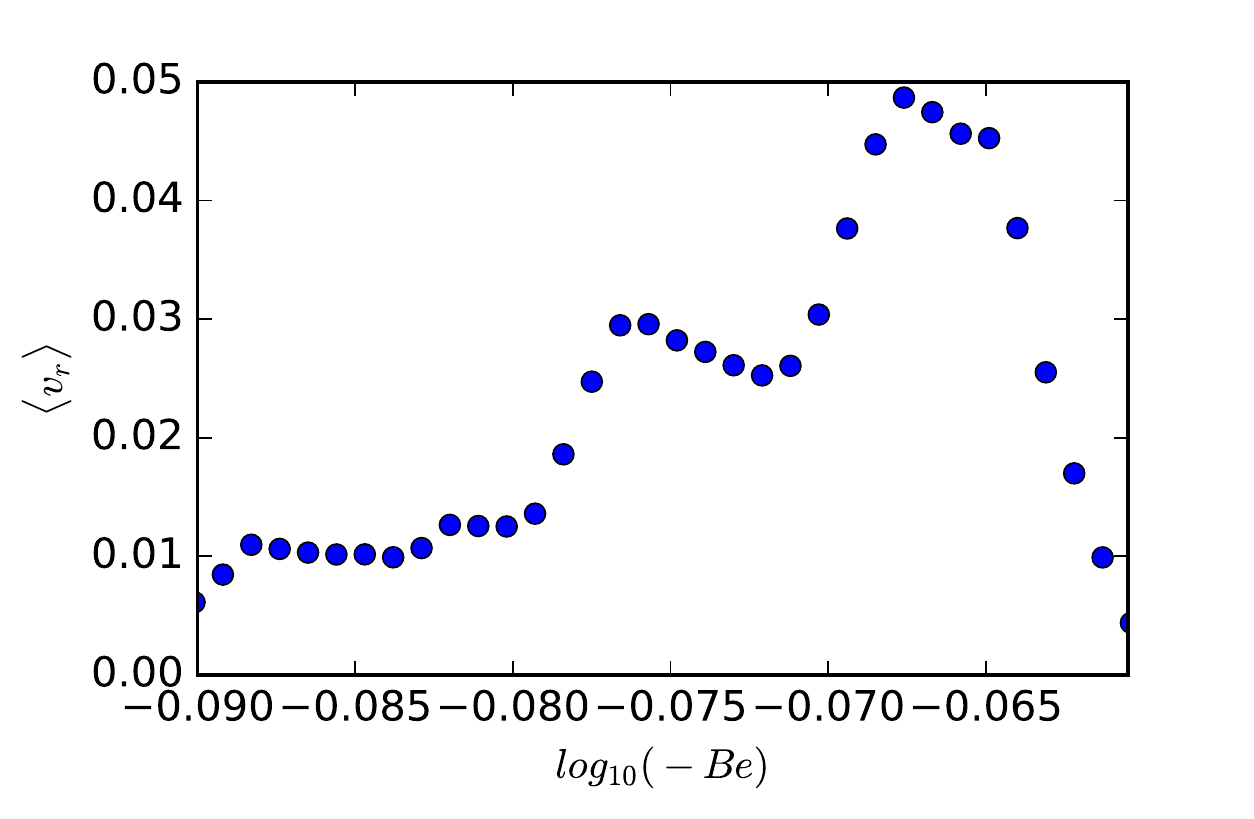}}
\caption{Histogram of the outflowing gas's $Be$ at $r=100\:r_g$ during the analysis portion of the simulation.  This is decomposed into the average density of the outflow as a function of $Be$ and the average velocity as a function of $Be$.
\label{fig-Bernoulli}}
\end{figure}

Shown in Figure \ref{fig-Bernoulli} is the distribution of outflowing mass fraction ($\dot{M}_{out}$), i.e. gas with $v_r>0$, at $r=100\:r_g$ as a function of $Be$.  On the surface, it looks like $\dot{M}_{out}$ is nearly independent of $Be$.  When we dig deeper into the distribution, we find that, in fact, there are two components to the outflow:  one component that is diffuse gas with high velocities and other that is dense gas with low radial velocity.  Also shown in Figure \ref{fig-Bernoulli} are the average density of the outflow and velocity as a function of $Be$.  When a bubble of hot gas is launched from the inner region of the disk, the gas has a large, positive radial velocity, but low density.  As it rises, higher density gas gets entrained in the wake of the bubble and pulled along to larger radius, although at a lower velocity.

\subsubsection{Synthetic Light Curves}

To finish our analysis, we use the cooling in our simulation to generate a synthetic light curve.  Since PLUTO conserves total energy in the simulation, the ``radiative losses" our cooling function imitates can be used as a proxy for the broadband emission.  There are obvious caveats to this exercise, namely that we use an \emph{ad hoc} cooling function to approximate the real, detailed radiation physics of the disk.  Regardless, it will provide a sense of what type of photometric variability could originate from this disk.

\begin{figure}
\includegraphics[width=0.5\textwidth]{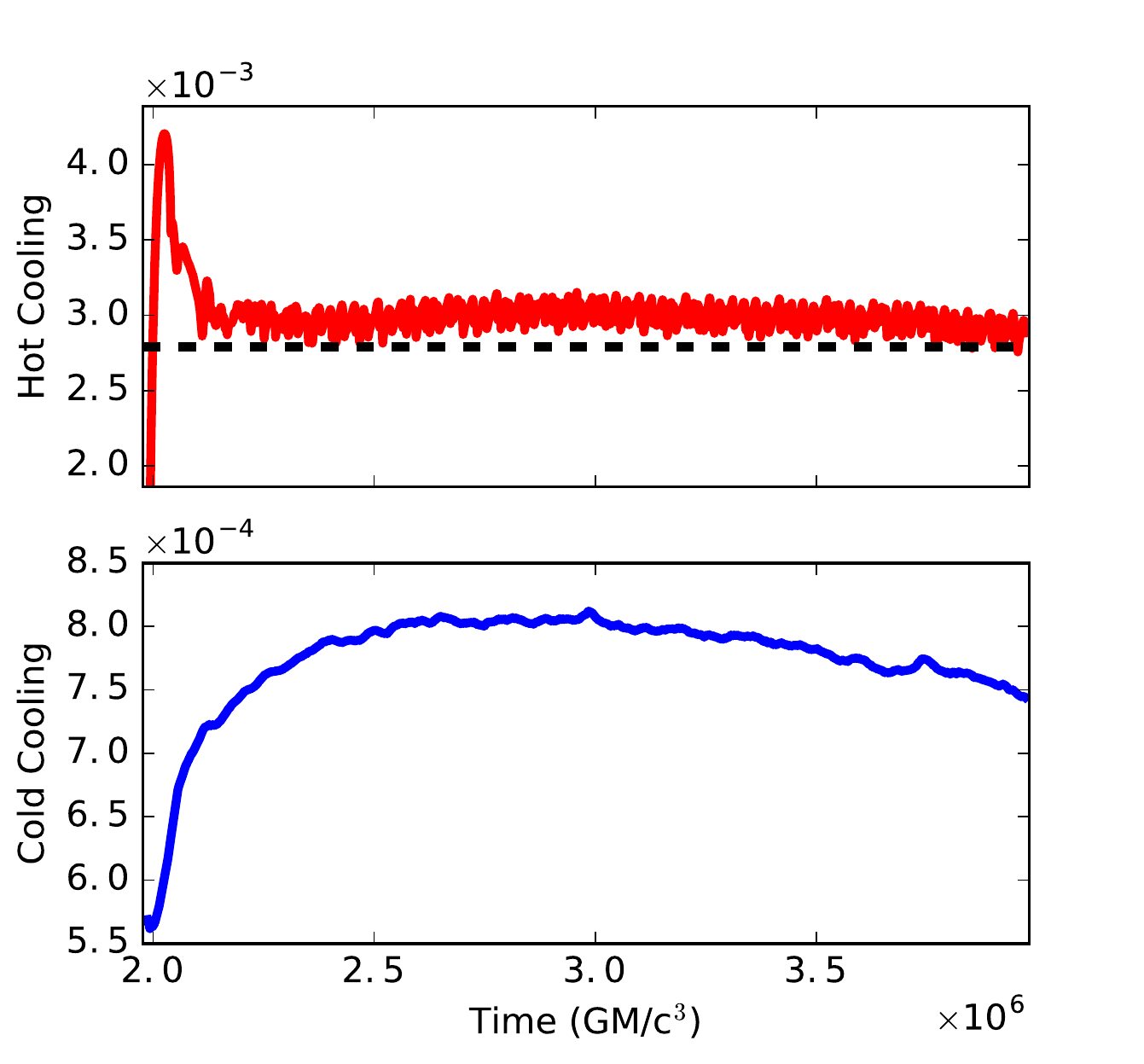}
\caption{Disk integrated synthetic light curve from the hot gas cooling ($L_h$, top panel) and cool gas cooling ($L_s$, blue line).  The cooling level during initialization of $L_p=2.8\times10^{-3}$ is shown as the black dashed line for reference. 
\label{fig-lightcurve}}
\end{figure}

Observationally, in astrophysical black hole systems, the higher energy emission is attributed to the hot, radiatively inefficient gas.  The lower energy, blackbody emission is attributed to the cooler, radiatively efficient, dense gas.  Following this paradigm, we use our previously defined temperature criteria to generate ``hard," $L_h$, and ``soft," $L_s$, band light curves, shown with the volume integrated cooling as a function of time of the hot and cold gas, respectively, in Figure \ref{fig-lightcurve}.

During the initialization, the large heating event causes an immediate spike in $L_h$ that decays down to a level of $L_h=3.0\times10^{-3}$ which it maintains for the rest of the simulation.  This is higher than the total pre-analysis emission level, $L_p$, of $2.8\times10^{-3}$.  $L_s$ is approximately a quarter less than the $L_h$, due in part to the fact that the hot gas fills a large portion of the simulation volume.  Interestingly, there is very little sign of the oscillatory behavior seen Figure \ref{fig-phase_frac} of the gas fraction at $r_{trans}$.  Instead, the light curves of each gas phase qualitatively resemble the the time trace of the respective mass accretion rates, shown in Figure \ref{fig-frac_mdot}.

\section{Discussion}
\label{sec-disc}

By introducing a bistable cooling law to a simple hydrodynamic viscous disk, we find that gas transitioning from the hot phase to the cold phase develops significant time variability which affects the evolution of the disk.  The transition between efficient and inefficient gas cooling leads to disk truncation and causes large, convective cells to develop in the inner region of the disk.  The presence of sub-Keplerian, outflowing gas in close proximity to the accretion disk generates a large $\theta-\phi$ viscous stress which has previously been neglected, but is influential here.  The inner, truncated region of the disk is occupied by hot gas that cools inefficiently.  Beyond the transition, the bulk of the gas resides in cool phase in a configuration similar to a standard Shakura $\&$ Sunyaev thin disk.  In general, the truncated disk in our simulation is qualitatively similar to the phenomenological model invoked to explain the spectral properties of BHBs in the low/hard and intermediate states like Cygnus X-1 \citep{1999MNRAS.305..457D},  XTE J1650-500 \citep{2006MNRAS.367..659D}, H1743-322, \citep{2009RAA.....9..901S}, MAXI J1543-564 \citep{2014MNRAS.440.2882R}, and GX 339-5 \citep{2015A&A...573A.120P}, which was expected by construction.  There are several notable differences between the simulation and the standard truncated disk model, which could manifest as observable signatures in BHB and LLAGN systems.

In the canonical HD truncated disk model, the inner accretion flow is assumed to be an ADAF and many of the observational results are interpreted with this assumption \citep[e.g.][]{1999ApJ...516..177G, 2000ApJ...534L..27S, 2000ApJ...541..120H, 2001ApJ...562L.133U, 2003ApJ...593..435M, 2005ApJ...620..905Y}.  Hence, it is interesting that a flow which closely resembles a CDAF naturally forms.  Even though there are qualitative similarities, the convective flow is ultimately distinct from the standard CDAF.  The primary way in which it differs is that the gas does not stay in the convective flow indefinitely as it is constantly cooling.  Once it is launched, the gas rises to larger radii, but then stalls and falls back.  The accretion rate is higher following the transition because of the interaction of the outflowing gas and the thin disk body.  It is, in fact, primarily the gas in the hot phase that accretes with very little cold gas making it to the inner boundary of the simulation.

One might hope the oscillatory gas fraction could be connected to observable variability, like a QPO, but that is not currently supported by our simulation.  QPOs are present in the intermediate states between the low-hard state and the high-soft states when the disk is assumed to be truncated, similar to what we modeled.  However, neither the mass accretion rate nor our simple cooling prescription show appreciable variability.  If the buoyant bubbles we capture in our simulation can be physically realized and related to QPO behavior, their signature must ultimately arise from more complex radiative processes than we have included.

Despite the obvious issues with connecting this HD accretion disk model to real black hole disks, it still plays an important role in providing a needed benchmark against which future MHD models of truncated disks can be compared, as we originally motivated.  In particular, the surprising complexity in the disk structure, angular momentum transport, and energetics demonstrate how poorly the behavior of disks in this state is understood, even in what should be a simple case.  The results from this simulation help to provide a bridge between the simple disk model currently used to understand state transitions in BHBs and LLAGNs and the more complex MHD physics that actually govern their evolution.

Of the many caveats and pitfalls associated with using an HD model to describe a real accretion disk, we will briefly discuss two.  The first concern is that the behavior is sensitive to the strength of the anomalous viscosity.  As the viscosity of the disk changes, the ability of the fluid to develop small scale structure and buoyantly rise also changes, which has been shown to affect the size and behavior of hydrodynamic, convective accretion flows \citep{1999MNRAS.303..309I}.  To test the sensitivity of this model we reran the full simulation with $\alpha=0.04$.  The development of this low $\alpha$-parameter was similar.  Granted, the amplitudes of the stress, $\dot{M}$, etc. changed because of the lower viscosity, but the distinguishing flow properties like the rapid inflow of gas along the skin of the disk, a single large convective eddy in the atmosphere, and the quasi-periodic development of buoyant bubbles at the truncation were all reproduced.  In all, we can be confident that the behavior we present in this paper is robust for reasonable $\alpha$-parameter values.

The biggest concern is the role MHD will play in altering the behavior of the disk.  One of the most conspicuous features of our simulation is the presence of a ``cold" stream of gas extending from the transition radius down to near the inner boundary.  Observationally, there is evidence that cool gas may be present near the black hole, even when the disk is thought to be truncated \citep{2002ApJ...577L..15M, 2002ApJ...578..348M}.  While it is tempting to make a connection between what develops in our simulation and what is empirically observed, this must be done cautiously.  In an astrophysical accretion disk the gas motions are dominated by vigorous, MHD turbulence with correlation lengths of order the disk scale height \citep{2009ApJ...694.1010G, 2011MNRAS.416..361B, 2016ApJ...826...40H}.  The stream of material in this simulation is smaller than that nominal scale, hence, it is reasonable to expect that even if this stream could form, it would quickly be disrupted by the turbulence and unable to exist in the way it does in our simulation.  Additionally, if this cold stream is destroyed by MRI driven turbulence, it should follow that the convective cells would disappear since they are launched from bubbles of gas trapped in the cold stream.  The absence of outflowing material against the disk would dramatically change the character from what we capture in this simulation since the vertical shear in the velocity field leads to the interesting behavior we observe.

Additionally, the properties of HD and MHD stresses are markedly different.  In this HD model the stresses, and subsequent angular momentum transport, originate from a true viscosity.  In an MHD disk, however, correlated radial and azimuthal fluctuations in both magnetic field and fluid velocity in anisotropic turbulence govern the angular momentum transport \citep{1994MNRAS.271..197B}.  This casts doubt on whether the vertical velocity shear plays an important role in the disk evolution and angular momentum transport.  Furthermore, the viscosity in this simulation does not vary.  In MHD, the stresses fluctuate on a dynamical time and are modulated by dynamo behavior on tens of dynamical times.  This natural variability in the effective viscosity could interact with the thermal instability and further manipulate the launching of hot bubbles of gas in the inner region of the disk.  An examination of this model within a MHD framework in Paper \RNum{2} will address these concerns and provide a rigorous comparison between the HD and MHD regimes.

\section{Conclusions}
\label{sec-conc}

Using a viscous, hydrodynamic simulation of an accretion disk with a bistable cooling function, we conduct a detailed analysis of the dynamics, angular momentum transport and energetics.  We find that a truncated configuration develops and that convective cells are launched from the evacuated region as cold gas flows inwards and undergoes delayed heating.  Pockets of hot gas form within the cold gas stream and buoyantly rise along the surface of the cooler disk which introduces a $\theta-\phi$ viscous stress component through a vertical shear between the coronal atmosphere and the disk itself.  This viscous stress component has been neglected in previous hydrodynamic treatments of accretion disks, but it plays in important role in the evolution of the disk by transferring additional angular momentum from the dense disk to the outflow and enhancing accretion in a thin layer long the surface of the disk.  In this truncated disk, the bulk of the accreted material is in the hot phase, rather than the cold phase, and the accretion rate is slightly elevated compared to that of the thin disk during the initialization.  The spontaneous transition of gas to the hot phase in the cold stream of material causes a quasi-periodic oscillation in the hot/cold gas fraction at the transition radius, but this does not translate in any appreciable fluctuations in the mass accretion rate, nor the volume integrated cooling ratio.  

With this baseline hydrodynamic model of a truncated disk, we can begin to understand more complex future MHD models.  Ultimately, the stresses we approximate as a hydrodynamic viscosity arise through the MRI, which behaves differently.  Additionally, the interesting behavior displayed by this simulation develops through purely HD processes and on small scales.  MHD turbulence should disrupt the flow enough that the global properties will be different; however, until a rigorous MHD test of a truncated accretion disk is completed, it is unknown how the disk evolution will be changed.

\acknowledgements

The authors thank the anonymous referee for useful comments that helped strengthen and clarify the paper.  CSR thanks support from a Sackler Fellowship (hosted by the Institute of Astronomy, Cambridge) and NASA under grant NNX15AC40G.  JDH thanks support by NASA under the NASA Earth and Space Science Fellowship program.  The authors acknowledge the University of Maryland supercomputing resources (http://www.it.umd.edu/hpcc) made available for conducting the research reported in this paper.


\begin{thebibliography}{}
\expandafter\ifx\csname natexlab\endcsname\relax\def\natexlab#1{#1}\fi

\bibitem[{{Armitage}(1998)}]{1998ApJ...501L.189A}
{Armitage}, P.~J. 1998, \apjl, 501, L189

\bibitem[{{Balbus} {et~al.}(1994){Balbus}, {Gammie}, \&
  {Hawley}}]{1994MNRAS.271..197B}
{Balbus}, S.~A., {Gammie}, C.~F., \& {Hawley}, J.~F. 1994, \mnras, 271,
  doi:10.1093/mnras/271.1.197

\bibitem[{{Balbus} \& {Hawley}(1991)}]{1991ApJ...376..214B}
{Balbus}, S.~A., \& {Hawley}, J.~F. 1991, \apj, 376, 214

\bibitem[{{Beckwith} {et~al.}(2011){Beckwith}, {Armitage}, \&
  {Simon}}]{2011MNRAS.416..361B}
{Beckwith}, K., {Armitage}, P.~J., \& {Simon}, J.~B. 2011, \mnras, 416, 361

\bibitem[{{Blandford} \& {Payne}(1982)}]{1982MNRAS.199..883B}
{Blandford}, R.~D., \& {Payne}, D.~G. 1982, \mnras, 199, 883

\bibitem[{{Das} \& {Sharma}(2013)}]{2013MNRAS.435.2431D}
{Das}, U., \& {Sharma}, P. 2013, \mnras, 435, 2431

\bibitem[{{Done} \& {Gierli{\'n}ski}(2006)}]{2006MNRAS.367..659D}
{Done}, C., \& {Gierli{\'n}ski}, M. 2006, \mnras, 367, 659

\bibitem[{{Done} {et~al.}(2007){Done}, {Gierli{\'n}ski}, \&
  {Kubota}}]{2007A&ARv..15....1D}
{Done}, C., {Gierli{\'n}ski}, M., \& {Kubota}, A. 2007, \aapr, 15, 1

\bibitem[{{Done} \& {Zycki}(1999)}]{1999MNRAS.305..457D}
{Done}, C., \& {Zycki}, P.~T. 1999, \mnras, 305, 457

\bibitem[{{Flock} {et~al.}(2011){Flock}, {Dzyurkevich}, {Klahr}, {Turner}, \&
  {Henning}}]{2011ApJ...735..122F}
{Flock}, M., {Dzyurkevich}, N., {Klahr}, H., {Turner}, N.~J., \& {Henning}, T.
  2011, \apj, 735, 122

\bibitem[{{Gammie} {et~al.}(1999){Gammie}, {Narayan}, \&
  {Blandford}}]{1999ApJ...516..177G}
{Gammie}, C.~F., {Narayan}, R., \& {Blandford}, R. 1999, \apj, 516, 177

\bibitem[{{Guan} {et~al.}(2009){Guan}, {Gammie}, {Simon}, \&
  {Johnson}}]{2009ApJ...694.1010G}
{Guan}, X., {Gammie}, C.~F., {Simon}, J.~B., \& {Johnson}, B.~M. 2009, \apj,
  694, 1010

\bibitem[{{Hawley}(2000)}]{2000ApJ...528..462H}
{Hawley}, J.~F. 2000, \apj, 528, 462

\bibitem[{{Hawley} \& {Balbus}(2002)}]{2002ApJ...573..738H}
{Hawley}, J.~F., \& {Balbus}, S.~A. 2002, \apj, 573, 738

\bibitem[{{Hawley} {et~al.}(2001){Hawley}, {Balbus}, \&
  {Stone}}]{2001ApJ...554L..49H}
{Hawley}, J.~F., {Balbus}, S.~A., \& {Stone}, J.~M. 2001, \apjl, 554, L49

\bibitem[{{Ho}(2008)}]{2008ARA&A..46..475H}
{Ho}, L.~C. 2008, \araa, 46, 475

\bibitem[{{Ho} {et~al.}(2000){Ho}, {Rudnick}, {Rix}, {Shields}, {McIntosh},
  {Filippenko}, {Sargent}, \& {Eracleous}}]{2000ApJ...541..120H}
{Ho}, L.~C., {Rudnick}, G., {Rix}, H.-W., {et~al.} 2000, \apj, 541, 120

\bibitem[{{Hogg} \& {Reynolds}(2016)}]{2016ApJ...826...40H}
{Hogg}, J.~D., \& {Reynolds}, C.~S. 2016, \apj, 826, 40

\bibitem[{{Homan} {et~al.}(2016){Homan}, {Neilsen}, {Allen}, {Chakrabarty},
  {Fender}, {Fridriksson}, {Remillard}, \& {Schulz}}]{2016ApJ...830L...5H}
{Homan}, J., {Neilsen}, J., {Allen}, J.~L., {et~al.} 2016, \apjl, 830, L5

\bibitem[{{Honma}(1996)}]{1996PASJ...48...77H}
{Honma}, F. 1996, \pasj, 48, 77

\bibitem[{{Igumenshchev} \& {Abramowicz}(1999)}]{1999MNRAS.303..309I}
{Igumenshchev}, I.~V., \& {Abramowicz}, M.~A. 1999, \mnras, 303, 309

\bibitem[{{Igumenshchev} \& {Abramowicz}(2000)}]{2000ApJS..130..463I}
---. 2000, \apjs, 130, 463

\bibitem[{{Igumenshchev} {et~al.}(2000){Igumenshchev}, {Abramowicz}, \&
  {Narayan}}]{2000ApJ...537L..27I}
{Igumenshchev}, I.~V., {Abramowicz}, M.~A., \& {Narayan}, R. 2000, \apjl, 537,
  L27

\bibitem[{{Igumenshchev} {et~al.}(2003){Igumenshchev}, {Narayan}, \&
  {Abramowicz}}]{2003ApJ...592.1042I}
{Igumenshchev}, I.~V., {Narayan}, R., \& {Abramowicz}, M.~A. 2003, \apj, 592,
  1042

\bibitem[{{Kippenhahn} \& {Thomas}(1982)}]{1982A&A...114...77K}
{Kippenhahn}, R., \& {Thomas}, H.-C. 1982, \aap, 114, 77

\bibitem[{{Kley} \& {Lin}(1992)}]{1992ApJ...397..600K}
{Kley}, W., \& {Lin}, D.~N.~C. 1992, \apj, 397, 600

\bibitem[{{Liu} {et~al.}(1999){Liu}, {Yuan}, {Meyer}, {Meyer-Hofmeister}, \&
  {Xie}}]{1999ApJ...527L..17L}
{Liu}, B.~F., {Yuan}, W., {Meyer}, F., {Meyer-Hofmeister}, E., \& {Xie}, G.~Z.
  1999, \apjl, 527, L17

\bibitem[{{Manmoto} \& {Kato}(2000)}]{2000ApJ...538..295M}
{Manmoto}, T., \& {Kato}, S. 2000, \apj, 538, 295

\bibitem[{{Manmoto} {et~al.}(2000){Manmoto}, {Kato}, {Nakamura}, \&
  {Narayan}}]{2000ApJ...529..127M}
{Manmoto}, T., {Kato}, S., {Nakamura}, K.~E., \& {Narayan}, R. 2000, \apj, 529,
  127

\bibitem[{{McClintock} {et~al.}(2003){McClintock}, {Narayan}, {Garcia},
  {Orosz}, {Remillard}, \& {Murray}}]{2003ApJ...593..435M}
{McClintock}, J.~E., {Narayan}, R., {Garcia}, M.~R., {et~al.} 2003, \apj, 593,
  435

\bibitem[{{McClintock} \& {Remillard}(2006)}]{2006csxs.book..157M}
{McClintock}, J.~E., \& {Remillard}, R.~A. 2006, {Black hole binaries}, ed.
  W.~H.~G. {Lewin} \& M.~{van der Klis}, 157--213

\bibitem[{{Meyer} \& {Meyer-Hofmeister}(1994)}]{1994A&A...288..175M}
{Meyer}, F., \& {Meyer-Hofmeister}, E. 1994, \aap, 288, 175

\bibitem[{{Mignone} {et~al.}(2007){Mignone}, {Bodo}, {Massaglia}, {Matsakos},
  {Tesileanu}, {Zanni}, \& {Ferrari}}]{2007ApJS..170..228M}
{Mignone}, A., {Bodo}, G., {Massaglia}, S., {et~al.} 2007, \apjs, 170, 228

\bibitem[{{Miller} {et~al.}(2002{\natexlab{a}}){Miller}, {Fabian}, {in't Zand},
  {Reynolds}, {Wijnands}, {Nowak}, \& {Lewin}}]{2002ApJ...577L..15M}
{Miller}, J.~M., {Fabian}, A.~C., {in't Zand}, J.~J.~M., {et~al.}
  2002{\natexlab{a}}, \apjl, 577, L15

\bibitem[{{Miller} {et~al.}(2002{\natexlab{b}}){Miller}, {Fabian}, {Wijnands},
  {Remillard}, {Wojdowski}, {Schulz}, {Di Matteo}, {Marshall}, {Canizares},
  {Pooley}, \& {Lewin}}]{2002ApJ...578..348M}
{Miller}, J.~M., {Fabian}, A.~C., {Wijnands}, R., {et~al.} 2002{\natexlab{b}},
  \apj, 578, 348

\bibitem[{{Mishra} {et~al.}(2016){Mishra}, {Fragile}, {Johnson}, \&
  {Klu{\'z}niak}}]{2016MNRAS.463.3437M}
{Mishra}, B., {Fragile}, P.~C., {Johnson}, L.~C., \& {Klu{\'z}niak}, W. 2016,
  \mnras, 463, 3437

\bibitem[{{Narayan} {et~al.}(2000){Narayan}, {Igumenshchev}, \&
  {Abramowicz}}]{2000ApJ...539..798N}
{Narayan}, R., {Igumenshchev}, I.~V., \& {Abramowicz}, M.~A. 2000, \apj, 539,
  798

\bibitem[{{Narayan} {et~al.}(1996){Narayan}, {McClintock}, \&
  {Yi}}]{1996ApJ...457..821N}
{Narayan}, R., {McClintock}, J.~E., \& {Yi}, I. 1996, \apj, 457, 821

\bibitem[{{Narayan} {et~al.}(2012){Narayan}, {S{\"A} dowski}, {Penna}, \&
  {Kulkarni}}]{2012MNRAS.426.3241N}
{Narayan}, R., {S{\"A} dowski}, A., {Penna}, R.~F., \& {Kulkarni}, A.~K. 2012,
  \mnras, 426, 3241

\bibitem[{{Narayan} \& {Yi}(1994)}]{1994ApJ...428L..13N}
{Narayan}, R., \& {Yi}, I. 1994, \apjl, 428, L13

\bibitem[{{Narayan} \& {Yi}(1995{\natexlab{a}})}]{1995ApJ...444..231N}
---. 1995{\natexlab{a}}, \apj, 444, 231

\bibitem[{{Narayan} \& {Yi}(1995{\natexlab{b}})}]{1995ApJ...452..710N}
---. 1995{\natexlab{b}}, \apj, 452, 710

\bibitem[{{Nemmen} {et~al.}(2014){Nemmen}, {Storchi-Bergmann}, \&
  {Eracleous}}]{2014MNRAS.438.2804N}
{Nemmen}, R.~S., {Storchi-Bergmann}, T., \& {Eracleous}, M. 2014, \mnras, 438,
  2804

\bibitem[{{Noble} {et~al.}(2009){Noble}, {Krolik}, \&
  {Hawley}}]{2009ApJ...692..411N}
{Noble}, S.~C., {Krolik}, J.~H., \& {Hawley}, J.~F. 2009, \apj, 692, 411

\bibitem[{{Plant} {et~al.}(2015){Plant}, {Fender}, {Ponti}, {Mu{\~n}oz-Darias},
  \& {Coriat}}]{2015A&A...573A.120P}
{Plant}, D.~S., {Fender}, R.~P., {Ponti}, G., {Mu{\~n}oz-Darias}, T., \&
  {Coriat}, M. 2015, \aap, 573, A120

\bibitem[{{Potter} \& {Balbus}(2014)}]{2014MNRAS.441..681P}
{Potter}, W.~J., \& {Balbus}, S.~A. 2014, \mnras, 441, 681

\bibitem[{{Potter} \& {Balbus}(2017)}]{2017arXiv170402485P}
---. 2017, ArXiv e-prints, arXiv:1704.02485

\bibitem[{{Proga} \& {Begelman}(2003)}]{2003ApJ...582...69P}
{Proga}, D., \& {Begelman}, M.~C. 2003, \apj, 582, 69

\bibitem[{{Proga} \& {Kallman}(2004)}]{2004ApJ...616..688P}
{Proga}, D., \& {Kallman}, T.~R. 2004, \apj, 616, 688

\bibitem[{{Proga} {et~al.}(2000){Proga}, {Stone}, \&
  {Kallman}}]{2000ApJ...543..686P}
{Proga}, D., {Stone}, J.~M., \& {Kallman}, T.~R. 2000, \apj, 543, 686

\bibitem[{{Quataert} \& {Gruzinov}(2000)}]{2000ApJ...539..809Q}
{Quataert}, E., \& {Gruzinov}, A. 2000, \apj, 539, 809

\bibitem[{{Rapisarda} {et~al.}(2014){Rapisarda}, {Ingram}, \& {van der
  Klis}}]{2014MNRAS.440.2882R}
{Rapisarda}, S., {Ingram}, A., \& {van der Klis}, M. 2014, \mnras, 440, 2882

\bibitem[{{Rees} {et~al.}(1982){Rees}, {Begelman}, {Blandford}, \&
  {Phinney}}]{1982Natur.295...17R}
{Rees}, M.~J., {Begelman}, M.~C., {Blandford}, R.~D., \& {Phinney}, E.~S. 1982,
  \nat, 295, 17

\bibitem[{{Remillard} \& {McClintock}(2006)}]{2006ARA&A..44...49R}
{Remillard}, R.~A., \& {McClintock}, J.~E. 2006, \araa, 44, 49

\bibitem[{{R{\'o}{\.z}a{\'N}ska} \& {Czerny}(2000)}]{2000A&A...360.1170R}
{R{\'o}{\.z}a{\'N}ska}, A., \& {Czerny}, B. 2000, \aap, 360, 1170

\bibitem[{{Rozyczka} {et~al.}(1994){Rozyczka}, {Bodenheimer}, \&
  {Bell}}]{1994ApJ...423..736R}
{Rozyczka}, M., {Bodenheimer}, P., \& {Bell}, K.~R. 1994, \apj, 423, 736

\bibitem[{{Ryu} \& {Goodman}(1992)}]{1992ApJ...388..438R}
{Ryu}, D., \& {Goodman}, J. 1992, \apj, 388, 438

\bibitem[{{Shakura} \& {Sunyaev}(1973)}]{1973A&A....24..337S}
{Shakura}, N.~I., \& {Sunyaev}, R.~A. 1973, \aap, 24, 337

\bibitem[{{Shields} {et~al.}(2000){Shields}, {Rix}, {McIntosh}, {Ho},
  {Rudnick}, {Filippenko}, {Sargent}, \& {Sarzi}}]{2000ApJ...534L..27S}
{Shields}, J.~C., {Rix}, H.-W., {McIntosh}, D.~H., {et~al.} 2000, \apjl, 534,
  L27

\bibitem[{{Sriram} {et~al.}(2009){Sriram}, {Agrawal}, \& {Raghurama
  Rao}}]{2009RAA.....9..901S}
{Sriram}, K., {Agrawal}, V.~K., \& {Raghurama Rao}, A. 2009, Research in
  Astronomy and Astrophysics, 9, 901

\bibitem[{{Stone} \& {Balbus}(1996)}]{1996ApJ...464..364S}
{Stone}, J.~M., \& {Balbus}, S.~A. 1996, \apj, 464, 364

\bibitem[{{Stone} \& {Pringle}(2001)}]{2001MNRAS.322..461S}
{Stone}, J.~M., \& {Pringle}, J.~E. 2001, \mnras, 322, 461

\bibitem[{{Stone} {et~al.}(1999){Stone}, {Pringle}, \&
  {Begelman}}]{1999MNRAS.310.1002S}
{Stone}, J.~M., {Pringle}, J.~E., \& {Begelman}, M.~C. 1999, \mnras, 310, 1002

\bibitem[{{Takahashi} {et~al.}(2016){Takahashi}, {Ohsuga}, {Kawashima}, \&
  {Sekiguchi}}]{2016ApJ...826...23T}
{Takahashi}, H.~R., {Ohsuga}, K., {Kawashima}, T., \& {Sekiguchi}, Y. 2016,
  \apj, 826, 23

\bibitem[{{Trump} {et~al.}(2011){Trump}, {Impey}, {Kelly}, {Civano}, {Gabor},
  {Diamond-Stanic}, {Merloni}, {Urry}, {Hao}, {Jahnke}, {Nagao}, {Taniguchi},
  {Koekemoer}, {Lanzuisi}, {Liu}, {Mainieri}, {Salvato}, \&
  {Scoville}}]{2011ApJ...733...60T}
{Trump}, J.~R., {Impey}, C.~D., {Kelly}, B.~C., {et~al.} 2011, \apj, 733, 60

\bibitem[{{Ulvestad} \& {Ho}(2001)}]{2001ApJ...562L.133U}
{Ulvestad}, J.~S., \& {Ho}, L.~C. 2001, \apjl, 562, L133

\bibitem[{{Urpin}(1984)}]{1984SvA....28...50U}
{Urpin}, V.~A. 1984, \sovast, 28, 50

\bibitem[{{Wu} {et~al.}(2016){Wu}, {Xie}, {Yuan}, \&
  {Gan}}]{2016MNRAS.459.1543W}
{Wu}, M.-C., {Xie}, F.-G., {Yuan}, Y.-F., \& {Gan}, Z. 2016, \mnras, 459, 1543

\bibitem[{{Yu} {et~al.}(2011){Yu}, {Yuan}, \& {Ho}}]{2011ApJ...726...87Y}
{Yu}, Z., {Yuan}, F., \& {Ho}, L.~C. 2011, \apj, 726, 87

\bibitem[{{Yuan} {et~al.}(2012){Yuan}, {Bu}, \& {Wu}}]{2012ApJ...761..130Y}
{Yuan}, F., {Bu}, D., \& {Wu}, M. 2012, \apj, 761, 130

\bibitem[{{Yuan} \& {Bu}(2010)}]{2010MNRAS.408.1051Y}
{Yuan}, F., \& {Bu}, D.-F. 2010, \mnras, 408, 1051

\bibitem[{{Yuan} {et~al.}(2005){Yuan}, {Cui}, \&
  {Narayan}}]{2005ApJ...620..905Y}
{Yuan}, F., {Cui}, W., \& {Narayan}, R. 2005, \apj, 620, 905

\bibitem[{{Yuan} {et~al.}(2015){Yuan}, {Gan}, {Narayan}, {Sadowski}, {Bu}, \&
  {Bai}}]{2015ApJ...804..101Y}
{Yuan}, F., {Gan}, Z., {Narayan}, R., {et~al.} 2015, \apj, 804, 101

\bibitem[{{Yuan} \& {Narayan}(2014)}]{2014ARA&A..52..529Y}
{Yuan}, F., \& {Narayan}, R. 2014, \araa, 52, 529

\bibitem[{{Zdziarski} \& {Gierli{\'n}ski}(2004)}]{2004PThPS.155...99Z}
{Zdziarski}, A.~A., \& {Gierli{\'n}ski}, M. 2004, Progress of Theoretical
  Physics Supplement, 155, 99

\end{thebibliography}
\end{document}